\documentclass[twocolumn,prb,superscriptaddress,showkeys,showpacs,notitlepage]{revtex4-1}
\pdfoutput=1
\usepackage{amsmath,amsfonts,amssymb,amsthm,epsfig,epstopdf,url,array}
\usepackage{dsfont}
\usepackage{bm}
\usepackage{dcolumn}
\usepackage{mathrsfs}
\usepackage{latexsym}
\usepackage{graphics}
\usepackage{textcomp}
\usepackage{mathrsfs}
\usepackage{indentfirst}
\usepackage{fancyhdr}
\usepackage{titletoc}
\usepackage{titlesec}
\usepackage{float}
\usepackage{ragged2e}
\usepackage{verbatim}
\usepackage{multirow}
\usepackage{mathtools}
\usepackage{bbold}
\usepackage{ragged2e}
\usepackage[hidelinks]{hyperref}
\usepackage{makecell}
\usepackage{color}
\usepackage[dvipsnames]{xcolor}

\begin{document}
\title{Singlet-Quintet Mixing in Spin-Orbit Coupled Superconductors with $j=3/2$ Fermions}
\author{Jiabin Yu}
\affiliation{Department of Physics, the Pennsylvania State University, University Park, PA, 16802}
\author{Chao-Xing Liu}
\email{cxl56@psu.edu}
\affiliation{Department of Physics, the Pennsylvania State University, University Park, PA, 16802}
\begin{abstract}
In non-centrosymmetric superconductors, spin-orbit coupling can induce
an unconventional superconducting state with a mixture
of s-wave spin-singlet and p-wave spin-triplet channels\cite{bauer2012non,PhysRevLett.87.037004,PhysRevLett.92.097001}, which leads to a variety of exotic phenomena,
including anisotropic upper critical field\cite{bauer2012non,yasuda2004superconducting,takeuchi2006specific,
settai2008huge,mukuda2010spin}, magnetoelectric effect\cite{bauer2012non,PhysRevB.65.144508,PhysRevLett.75.2004,PhysRevB.72.024515}, topological superconductivity\cite{PhysRevB.79.094504,PhysRevB.79.060505},
{\it et al}\cite{bauer2012non}.
It is commonly thought that inversion symmetry breaking is substantial
for pairing-mixed superconducting states.
In this work, we theoretically propose that a new type of pairing-mixed
state, namely the mixture of s-wave spin-singlet and d-wave spin-quintet
channels, can be induced by spin-orbit coupling even in the presence of inversion symmetry
when electrons effectively carry ``spin-3/2" in superconductors. As a physical consequence of the singlet-quintet pairing mixing, topological nodal-line superconductivity is found in such system
and gives rise to flat surface Majorana bands. Our work provides a possible explanation of unconventional superconducting behaviors observed in superconducting half-Heusler compounds\cite{PhysRevLett.116.177001,PhysRevB.84.220504,kim2016beyond,PhysRevB.86.064515,PhysRevLett.116.137001}. 	
\end{abstract}
\maketitle

In the Bardeen-Cooper-Schrieffer theory, the s-wave spin-singlet pairing
relies on the presence of both time reversal and inversion symmetry in superconductors (SCs).
In non-centrosymmetric SCs, the absence of inversion symmetry can give rise to anti-symmetric spin-orbit coupling (SOC) with odd parity, and results in a mixture of s-wave spin-singlet (even parity) and p-wave spin-triplet (odd parity) pairings \cite{bauer2012non,PhysRevLett.87.037004,PhysRevLett.92.097001}.
Due to the opposite parities of singlet and triplet pairings,
only anti-symmetric SOC is considered in pairing mixing mechanism \cite{bauer2012non}, while symmetric SOC with even parity is normally overlooked in non-centrosymmetric SCs.
However, we will show below this is not true if electrons carry ``spin-3/2''.
Here ``spin'' refers to total angular momentum $j$,
which is a combination of 1/2-spin and angular momentum of p atomic orbitals ($l=1$),
of basis electronic states.
Such superconductivity with $j=3/2$ electrons was recently proposed in superconducting
half-Heusler compounds \cite{PhysRevLett.116.177001},
where unconventional superconducting behaviors,
including low carrier density
\cite{PhysRevB.84.220504,kim2016beyond,PhysRevB.86.064515,PhysRevLett.116.137001},
power-law temperature dependence of London penetration depth \cite{kim2016beyond}
and large upper critical field\cite{PhysRevB.86.064515}, have been observed.
Superconductivity with spin-3/2 fermions has also been considered in cold atom systems\cite{wu2006hidden}.
In contrast to spin-1/2 SCs with only singlet and triplet states,
the Cooper pairs of $j=3/2$ electrons can carry total spin $S=0$ (singlet), 1 (triplet), 2 (quintet) and 3 (septet).
In this work, we demonstrate a new pairing-mixed state, namely the mixing
between s-wave spin-singlet and d-wave spin-quintet pairings,
can appear in spin-orbit coupled SCs with $j=3/2$ electrons, even in the presence of inversion symmetry.
In particular, we will illustrate the role of symmetric SOC (parity-even) in
the singlet-quintet mixing and how such pairing mixing can give rise to topological
nodal-line superconductivity (TNLS).

We start from electronic band structures of half-Heusler compounds and illustrate
the origin of $j=3/2$ electrons.
The energy bands near the Fermi energy in half-Heusler compounds
are s-type bands ($\Gamma_6$ bands) and p-type bands, where the latter is split into
$j=3/2$ bands ($\Gamma_8$ bands) and $j=1/2$ bands ($\Gamma_7$ bands) by SOC \cite{winkler2003spin}.
For half-Heusler SCs with p-type of carriers like YPtBi\cite{PhysRevB.84.220504},
only the $\Gamma_8$ bands are relevant\cite{chadov2010tunable},
and can be described by four-component wavefunctions, labeled as $|j,j_z\rangle$,
with total angular momentum $j=3/2$ that can be effectively regarded as ``spin''
and  $j_z=\pm 1/2, \pm 3/2$.
The low energy physics of the $\Gamma_8$ bands is described by the so-called
Luttinger model\cite{chadov2010tunable,PhysRev.102.1030} with the Hamiltonian
\begin{equation}\label{Eqn:h}
h(\mathbf{k})
=
\xi_{\mathbf{k}}\Gamma^0
+h_{SOC}(\mathbf{k})=\xi_{\mathbf{k}}\Gamma^0+
c_1 \sum_{i=1}^{3}g_{\mathbf{k},i}\Gamma^i+c_2 \sum_{i=4}^{5}g_{\mathbf{k},i}\Gamma^i,
\end{equation}
on the basis wavefunctions of $|j,j_z\rangle$,
where $\xi_{\mathbf{k}}=\frac{1}{2m}k^2-\mu$ with the chemical potential $\mu$.
The detailed forms of five d-orbital cubic harmonics $g_i$'s and six
4-by-4 matrices $\Gamma^i$ ($i=0,\dots,5$)
are defined in Sec.A of supplementary materials (SMs).
The above Hamiltonian only includes symmetric SOC term $h_{SOC}$,
while the antisymmetric SOC that breaks inversion will be discussed in the end.
The Luttinger Hamiltonian $h(\mathbf{k})$ is $O(3)$ invariant if $c_1=c_2$, and its symmetry is reduced
to $O_h$ group if $c_1\neq c_2$.
The eigen-states of $h(\mathbf{k})$ are doubly degnerate with eigen-energies
$\xi_{\pm}(\mathbf{k})=k^2/(2m_{\pm})-\mu$, where the subscript $\pm$ labels two spin-split bands,
and $m_{\pm}=m/(1\pm 2mQ_c)$ with $Q_c=\sqrt{c_1^2 Q_1^2+c_2^2 Q_2^2}$, $Q_1=\sqrt{\hat{g}^2_{1}+\hat{g}^2_{2}+\hat{g}^2_{3}}$, $Q_2=\sqrt{\hat{g}^2_{4}+\hat{g}^2_{5}}$ and $\hat{g}_i=g_i/k^2$.
We focus on the parameter regime with $m<0$ \cite{PhysRevLett.119.136401},
$\mu<0$ (p-type carriers), and $c_1 c_2 >0$ for simplicity.
With the choice of these parameters, the effective mass $m_-$ of the $\xi_-$ band is always negative
while there are three different regimes for $m_+$ of the $\xi_+$ band: (I) $m_+<0$, (II) $m_+>0$,
and (III) the sign of $m_+$ being angular dependent.
Energy dispersions and Fermi surface shapes in these three regimes
are depicted in Fig.1a.
In realistic materials, the regime I appears for the normal band structure when
$\Gamma_6$ bands have higher energy than $\Gamma_8$ bands while the regime II exists
for the inverted band structure with $\Gamma_6$ bands below $\Gamma_8$ bands.\cite{chadov2010tunable}
In the regime III, the $\xi_+$ band disperses oppositely along the direction $\Gamma-X$
and $\Gamma-L$, thus forming a saddle point at $\Gamma$ (Fig.1a(iii))
and hyperbolic Fermi surface (Fig.1a(vi)).
We notice that in realistic materials\cite{PhysRevLett.116.137001, PhysRevLett.119.136401},
the $\xi_+$ bands should eventually bend up at a large momentum
in all directions (the dashed lines in Fig. 1a(iii) and (vi)).
Thus, the Luttinger model is only valid in a small momentum region around $\Gamma$ in the regime III.

Next we will discuss the interaction Hamiltonian and the possible superconducting pairings in the Luttinger model.
Several types of pairing forms have been discussed in literature, including
mixed singlet-septet pairing\cite{PhysRevLett.116.177001,kim2016beyond,PhysRevB.96.144514,PhysRevB.96.094526},
s-wave quintet pairing\cite{PhysRevLett.116.177001,roy2017topological,PhysRevB.96.094526,boettcher2017unconventional}
, d-wave quintet pairing\cite{PhysRevLett.117.075301,venderbos2017pairing}
, odd-parity (triplet and septet) parings\cite{PhysRevLett.117.075301,venderbos2017pairing,savary2017superconductivity}, {\it et al}\cite{venderbos2017pairing}.
In particular, it is argued that s-wave singlet can be mixed with p-wave septet due
to antisymmetric SOC \cite{PhysRevLett.116.177001,kim2016beyond}.
Here we focus on possible pairing mixing induced by symmetric SOC $h_{SOC}$.
In analog to the singlet-triplet mixing, in which
the p-wave character of triplet channel originates from the p-wave nature of anti-symmetric
SOC term\cite{PhysRevLett.92.097001},
it is natural to expect that the pairing channel that is mixed into singlet channel
due to $h_{SOC}$ should have d-wave nature with orbital angular momentum $L=2$,
given the d-wave $g_{ {\bf k},i }$ in $h_{SOC}$.
According to the symmetry classification of gap functions for $j=3/2$ fermions\cite{savary2017superconductivity} and the coupled linearized gap equations (See Sec.B4 of SMs),
the only channel that can be mixed with s-wave singlet channel
is d-wave quintet channel, which carries $(L,S,J)$=(2,2,0) with spin $S$=2 (quintet)
and total angular momentum $J$=0 (${\bf J=L+S}$) for the Cooper pair, under $O(3)$ symmetry.
Here we focus on a minimal $O(3)$-invariant interaction
\begin{equation}\label{Eqn:HI}
H_I=\frac{1}{2\mathcal{V}}
\left(
V_0 P_s P_s^{\dagger}
+
V_1 P_q P_q^{\dagger}
\right)
\end{equation}
in the s-wave singlet and d-wave quintet channels,
where $P_s=\sum_{\mathbf{k}}c^{\dagger}_{\mathbf{k}}(\Gamma^0 \gamma/2)(c^{\dagger}_{-\mathbf{k}})^T$,
$P_q=\sum_{\mathbf{k}}c^{\dagger}_{\mathbf{k}}(a^2 \mathbf{g}_{\mathbf{k}}\cdot\mathbf{\Gamma}
\gamma/2)(c^{\dagger}_{-\mathbf{k}})^T$, and $V_0$ and $V_1$
stand for the s-wave and d-wave interaction parameters, respectively.
Here $c^{\dagger}_{\mathbf{k}}$ is the four-component creation operator on the basis $|j,j_z\rangle$,
$\gamma=-\Gamma^1\Gamma^3$ is the time-reversal matrix,
$\mathcal{V}$ is volume and $a$ is lattice constant.
As discussed in Sec.B5 of SMs, the above interaction Hamiltonian $H_I$ can be extracted
from the electron-optical phonon interaction proposed in Ref.\cite{savary2017superconductivity}.

According to the interaction in Eq.\ref{Eqn:HI}, the gap function should take the form
$\Delta(\mathbf{k})=\Delta_0 (\Gamma^0 \gamma/2)$ $+\Delta_1 (a^2 \mathbf{g}_{\mathbf{k}}\cdot\mathbf{\Gamma} \gamma/2)$,
in which $\Delta_0$ and $\Delta_1$ represent s-wave singlet and d-wave quintet channels, respectively.
The corresponding coupled linearized gap equation can be derived as
(Sec.B6 of SMs)
\begin{equation}
\label{lnearized_gap_equation}
\left(
\begin{array}{c}
\tilde{\Delta}_{0}\\
\tilde{\Delta}_{1}
\end{array}
\right)
=
x\left(
\begin{array}{cc}
\frac{1}{2}\lambda_{0}y_1 & \frac{1}{2}\lambda_{0} y_2\\
\frac{1}{2}\tilde{\lambda}_{1} y_2& \frac{1}{2}\tilde{\lambda}_{1} y_3
\end{array}
\right)
\left(
\begin{array}{c}
\tilde{\Delta}_{0}\\
\tilde{\Delta}_{1}
\end{array}
\right),
\end{equation}
where $x=\ln[2e^{\bar{\gamma}}\epsilon_c/(\pi k_BT)]$, $\bar{\gamma}$ is the Euler constant,
$k_B$ is Boltzman constant, $T$ is the critical temperature,
$\epsilon_c$ is the energy cut-off for the attractive interaction($V_{0,1}<0$),
$\lambda_0= -V_0 N_0$ and $\tilde{\lambda}_1= -(2m \mu a^2)V_1 N_0$ are the normalized
interaction parameters with the density of state $N_0$,
and $\tilde{\Delta}_{0}$ $=$ $\Delta_{0}\text{sgn}(c_1)$ and
$\tilde{\Delta}_{1}$ $=$ $\Delta_{1}(2m \mu a^2)$ are the normalized order parameters.
The band information is included in the functions $y_{1,2,3}$.
In the limit $\epsilon_c/2 Q_c k_F^2\ll 1$, $k_B T/\epsilon_c\ll 1$ and $\epsilon_c/|\mu| \ll 1$,
the functions $y_{1,2,3}$ can be perturbatively expanded as
$y_1$ $=$ $\langle\text{Re}[\tilde{m}_{-}^{3/2}+\tilde{m}_{+}^{3/2}]\rangle$,
$y_2$ $=$ $\langle \text{Re}[-\tilde{m}_{-}^{5/2}+\tilde{m}_{+}^{5/2}]$ $f_Q\rangle$
and
$y_3$ $=$ $\langle \text{Re}[\tilde{m}_{-}^{7/2}+\tilde{m}_{+}^{7/2}]$ $f_Q^2\rangle$
up to the leading order, where $\text{Re}[...]$ means taking the real part, $\langle ...\rangle$ represents averaging over the solid angle
,$f_Q=(|c_1|Q_1^2+ |c_2|Q_2^2)/Q_c$
and $\tilde{m}_{\pm}=m_{\pm}/m$
are the normalized effective masses of the $\xi_{\pm}$ bands.
As demonstrated in Sec.B6 of SMs, zero $c_{1,2}$
can lead to a vanishing off-diagonal term in the gap equation ($y_2=0$) due to $\tilde{m}_{+}=\tilde{m}_{-}$,
thus revealing the essential role of $h_{SOC}$ in singlet-quintet mixing.

\begin{figure}[t]
\includegraphics[width=\columnwidth]{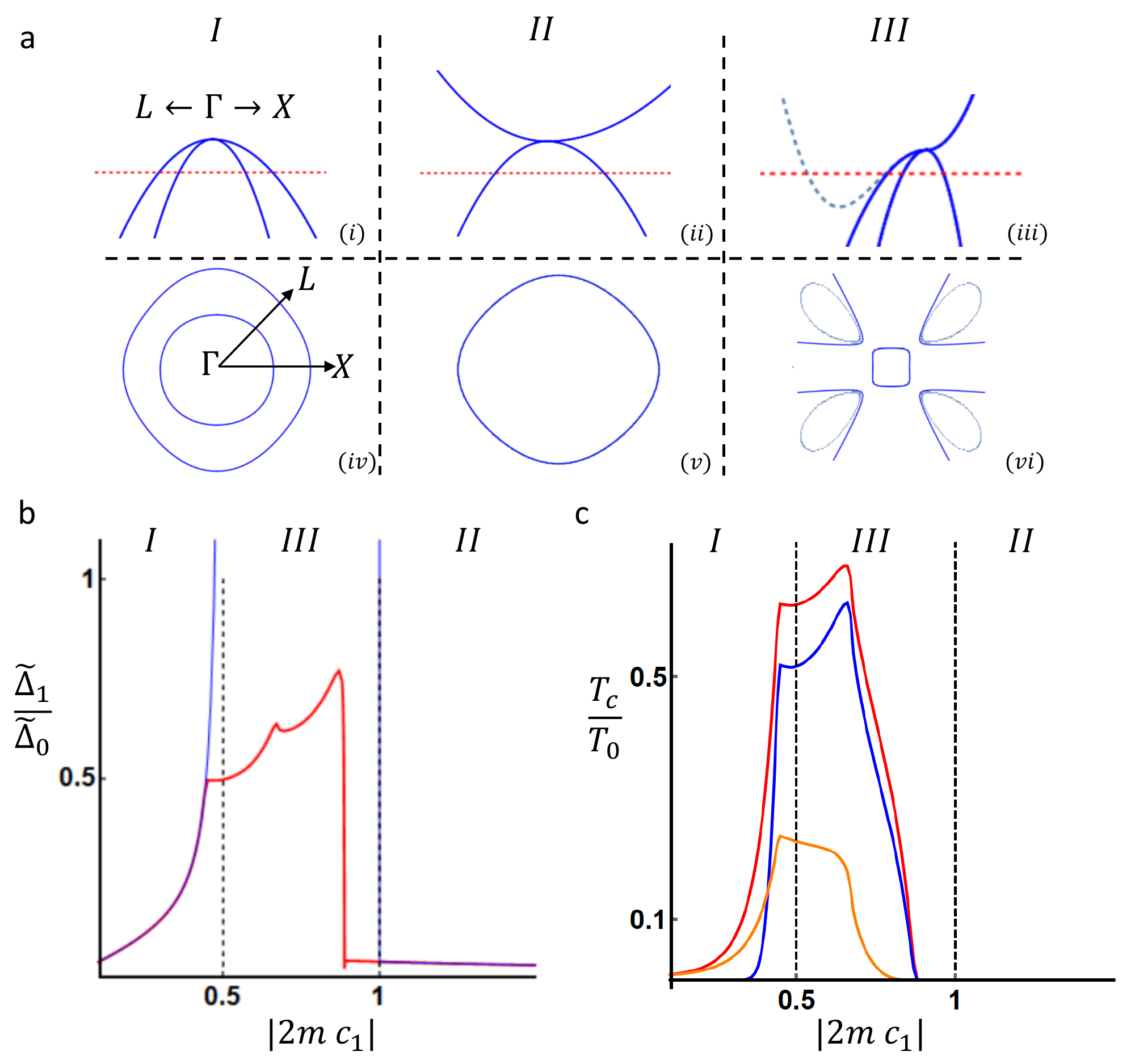}
\caption{
(a) Energy dispersions along $X\leftarrow\Gamma\rightarrow L$ are shown in (i), (ii) and (iii) (Solid lines), and the corresponding Fermi surfaces in $X-\Gamma-L$ plane are shown in (iv), (v) and (vi) for the Luttinger model in the regime I, II and III, respectively. The dashed lines in (iii) and (vi) depict energy dispersions and Fermi surfaces for the regime III in realistic compounds. The red dashed line represents the chemical potential.
The ratio $\tilde{\Delta}_1/\tilde{\Delta}_0$ and the critical temperature $T_c$ are shown in (b) and (c) as a function of $|2mc_1|$ for $c_2=2 c_1$, $\lambda_0=0.2$ $\tilde{\lambda}_1=0.1\lambda_0$ and $T_0=2e^{\bar{\gamma}}\epsilon_c/(\pi k_B)$. The blue and red lines in (b) corresponds to the case without and with momentum cut-off $\Lambda=3\sqrt{2m\mu}$, respectively.
The red line in (c) stands for the critical temperature with pairing mixing while the blue and orange lines give the critical temperatures of pure quintet and singlet channels without mixing, respectively.}
\end{figure}

By solving Eq. (\ref{lnearized_gap_equation}),
the mixing ratio $\tilde{\Delta}_{1}/\tilde{\Delta}_{0}$
is evaluated numerically as a function of $|2 m c_1|$ in Fig.1b (blue line)
for $c_2=2c_1$ and $\tilde{\lambda}_1=0.1\lambda_0$,
which reveals different behaviors in three parameter regimes I, II and III.
$\tilde{\Delta}_{1}/\tilde{\Delta}_{0}$ increases rapidly with $|2 m c_1|$ in regime I,
and diverges in regime III.
The dominant d-wave quintet pairing in regime III originates from the faster divergence of
$y_3$ compared to $y_{1,2}$ in Eq. (\ref{lnearized_gap_equation}).
To take into account the limitation of the Luttinger model in parameter regime III,
a momentum cut-off $\Lambda$
is introduced in computing $y_{1,2,3}$ as shown in Sec.B7 of SMs. With $\Lambda$, a peak strucure of
$\tilde{\Delta}_{1}/\tilde{\Delta}_{0}$ (the red line in Fig. 1b) is found and
 confirms the dominant role of d-wave quintet pairing in regime III.
Other features of $\tilde{\Delta}_{1}/\tilde{\Delta}_{0}$ in the regime III
(e.g. the kinks) are discussed in Sec.B7
of SMs. With further increasing $|2 m c_1|$ (regime II),
$\tilde{\Delta}_{1}/\tilde{\Delta}_{0}$ drops rapidly due to the disappearance of Fermi surface
for the $\xi_+$ bands and thus simple s-wave singlet pairing dominates in this regime.
In Fig.1c, the critical temperatures $T_c$ as a function of $|2 m c_1|$ are revealed by the red line for the pairing
mixing case, and by the orange and blue lines for the pure singlet and quintet cases, respectively.
We find that (1) pairing mixing can help enhance critical temperature;
and (2) singlet pairing dominates for most of regime I and the entire regime II while quintet pairing
plays a vital role around regime III.

Similar to the singlet-triplet mixing in non-centrosymmetric SCs\cite{bauer2012non,PhysRevB.79.094504,PhysRevB.85.024522,PhysRevB.84.020501,PhysRevB.83.064505},
a physical consequence of singlet-quintet mixing is
the existence of TNLS in certain parameter regimes.
The topological property of superconducting phases can be extracted from the Bogoliubov-de Gennes Hamiltonian
with the gap function determined by the gap equation (Eq. \ref{lnearized_gap_equation}).
TNLS can exist in the regime II when $V_0<0$ and $V_1>0$
and in the regime I and III as long as $V_0<0$ (Sec.C 2, 3, 5 and 7 of SMs).
Here we focus on the regime I with normal band structure and $V_{0,1}<0$.
Fig.2a shows the phase diagram in the parameter space spanned by SOC strength $|2 m c_1|$ and interaction strength ratio
$\tilde{\lambda}_1/\lambda_0$.
Nodal rings are found in the yellow and red regions of Fig.2a for the $\xi_-$ band (Fig. 2b and e).
Due to time reversal and inversion, a four-fold degeneracy exists at each point on the nodal ring.
Fig. 2b (i-iv) reveals the evolution of nodal rings along the path $\alpha$ depicted in the inset of Fig. 2a.
Six nodal rings first emerge and center around the $(001)$, $(010)$
and $(100)$ axes in Fig.2b (i).
These nodal rings expand (Fig.2b (ii)) and touch each other,
resulting in a Lifshitz transition (Fig.2b (iii)).
After the transition, eight nodal rings with their centers at the $(111)$
and other three equivalent axes (Fig.2b (iv)) shrink to eight points and eventually disappear.
Topological nature of these nodal rings can be extracted by evaluating
topological invariant $N_w$ of one dimensional AIII class \cite{PhysRevB.84.060504}
along the loop shown by the red circle in Fig.2b(i) (See Sec.C4 of SMs for detals).
Direct calculation gives $N_{w}=\pm 2$, coinciding with four-fold degeneracy of the nodal rings.
Non-zero $N_w$ also implies the existence of Majorana flat bands at the surface of TNLS.
Fig. 2c(More details in Sec.C8 of SMs) and d show the zero-energy density of states and the energy dispersions at the (111)
surface, which are calculated from the iterative Green function method \cite{sancho1985highly}.
The evolution of surface Majorana flat bands follows that of nodal ring structures
(see Fig. 2c (i-iv) and d (i-iv)).
Additional nodal rings exist in the red region of the phase diagram (Fig. 2a), as shown in Fig. 2e.

\begin{figure}[t]
\includegraphics[width=1\columnwidth]{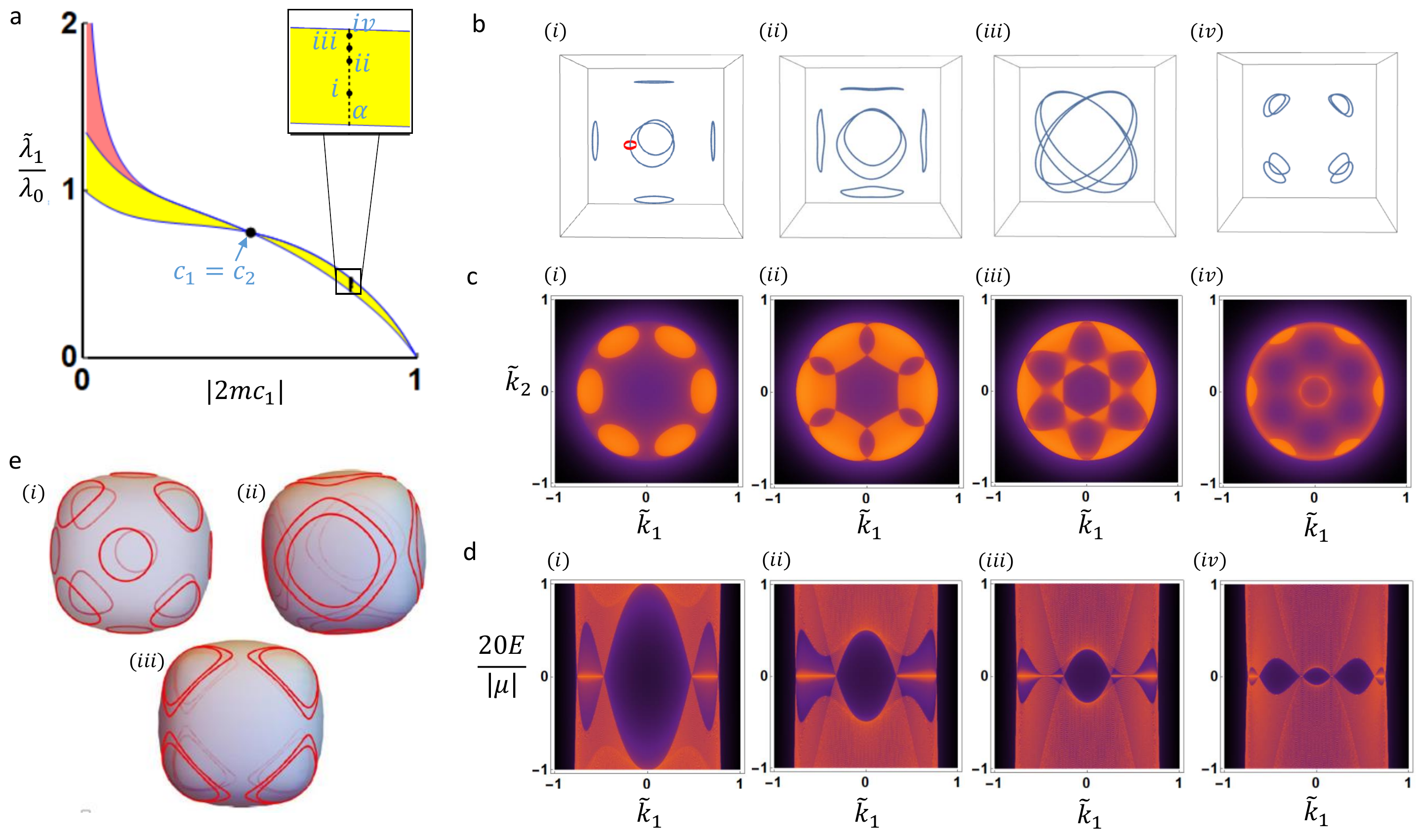}
\caption{
(a) shows the phase diagram in the parameter space spanned by interaction strength ratio $\tilde{\lambda}_1/\lambda_0$ and symmetric SOC strength $|2m c_1|$. In the yellow and red regions, the system are nodal. In the inset, the dashed line indicates the path $\alpha$ ($2 m |c_1|=-0.8$) with four points $i,...,iv$ on it.
Here $\tilde{\lambda}_1/\lambda_0= 0.4246, 0.4507, 0.4615, 0.4716$ for $(i)$, $(ii)$, $(iii)$, $(iv)$, respectively.
(b),(c) and (d) show the bulk nodal line structures (blue lines), zero-energy density of states on (111) surface and energy dispersion along $(11\bar{2})$ axis on (111) surface for the four points $i,...,iv$ in the inset of (a).
The red circle in (i) of (b) shows a typical path along which the topological invariant is calculated.
$\tilde{k}_{1,2}=k_{1,2}/\sqrt{2m\mu}$ are momenta along $(11\bar{2})$ and $(\bar{1}10)$, respectively, and $c_1>0$ and $\tilde{\Delta}_0/|\mu|=1$ are chosen.
(e) shows three typical nodal structures in the red region of (a).
Parameters are chosen as $2m|c_1|=-0.12$, $2m|c_2|=-0.5$ and $\tilde{\lambda}_1/\lambda_0=1.12$ for (i), $2m|c_1|=-0.12$, $2m|c_2|=-0.5$ and $\tilde{\lambda}_1/\lambda_0=1.155$ for (ii), and $2m|c_1|=-0.08$, $2m|c_2|=-0.5$ and $\tilde{\lambda}_1/\lambda_0=1.329$ for (iii).}
\end{figure}

We finally discuss the experimental implications of our theory.
Previous theoretical studies on half-Heusler SCs mainly focus on the compounds in regime II (inverted band structure),
while our study suggests that regimes I (normal band structure)
and III (a special case of inverted band structure) are more interesting due to strong singlet-quintet mixing.
Superconductivity has been found in DyPdBi and YPdBi with normal band structure \cite{nakajima2015topological}
and critical temperatures around $0.8K$ and $1.6K$, respectively, thus providing good candidates for TNLS.
YPtBi is a SC with inverted band
structure\cite{chadov2010tunable} and recent first principles calculations
\cite{PhysRevLett.116.137001,PhysRevLett.116.177001,PhysRevLett.119.136401} suggest that its energy dispersion
might belong to regime III, although debates still exist\cite{PhysRevLett.116.177001,kim2016beyond}.
Evidence of TNLS has been found in the penetration depth experiment\cite{kim2016beyond} .
Previous study attributes the nodal structure to the p-wave septet pairing mixed with
subdominant s-wave singlet pairing due to asymmetric SOC\cite{PhysRevLett.116.177001,kim2016beyond}.
Our work here provides an alternative explanation of the nodal structure as a result of
singlet-quintet mixing induced by symmetric SOC $h_{SOC}$.
In realistic half-Heusler compounds, the energy scale of symmetric SOC ($\sim 1eV$) is
two orders of magnitude larger than anti-symmetric SOC ($\sim 0.01eV$)
\cite{PhysRevLett.116.177001,savary2017superconductivity}. Thus, anti-symmetric SOC should be regarded as a perturbation and its influence on nodal-ring structures is discussed in Sec.C6 of SMs.
Furthermore, the interaction in s-wave singlet channel is normally the dominant mechanism for superconductivity
in weakly correlated materials.
Therefore, we expect singlet-quintet mixing should be dominant over singlet-septet mixing
and response for the nodal line structure in realistic SCs.
Our new pairing mixing mechanism opens up a door to explore other exotic superconducting
phenomena in spin-orbit coupled SCs with $j=3/2$ electrons.

\section*{Acknowledgment} 
JY owes a large amount of thanks to Lun-Hui Hu for patiently answering his questions on superconductivity. JY also thanks Rui-Xing Zhang, Yang Ge and Jian-Xiao Zhang for helpful discussion. CXL and JY acknowledge the support from Office of Naval Research (Grant No. N00014-15-1-2675).

\bibliography{Three_Half_Paring}

\begin{widetext}
\appendix
\section{Expressions and Conventions}
The five d-orbital cubic harmonics are given by \cite{PhysRevB.69.235206}
\begin{equation}
\left\{
\begin{array}{l}
g_{\mathbf{k},1}=\sqrt{3} k_y k_z\\
g_{\mathbf{k},2}=\sqrt{3} k_z k_x\\
g_{\mathbf{k},3}=\sqrt{3} k_x k_y\\
g_{\mathbf{k},4}=\frac{\sqrt{3}}{2} (k_x^2-k_y^2)\\
g_{\mathbf{k},5}=\frac{1}{2}(2 k_z^2-k_x^2-k_y^2)\\
\end{array}
\right..
\end{equation}

The angular momentum matrices of $J=\frac{3}{2}$ are written as \cite{PhysRevB.69.235206}
\begin{equation}
J_x=\left(
\begin{array}{cccc}
 0 & \frac{\sqrt{3}}{2} & 0 & 0 \\
 \frac{\sqrt{3}}{2} & 0 & 1 & 0 \\
 0 & 1 & 0 & \frac{\sqrt{3}}{2} \\
 0 & 0 & \frac{\sqrt{3}}{2} & 0 \\
\end{array}
\right)
\end{equation}
\begin{equation}
J_y=\left(
\begin{array}{cccc}
 0 & -\frac{i \sqrt{3}}{2} & 0 & 0 \\
 \frac{i \sqrt{3}}{2} & 0 & -i & 0 \\
 0 & i & 0 & -\frac{i \sqrt{3}}{2}  \\
 0 & 0 & \frac{i \sqrt{3}}{2} & 0 \\
\end{array}
\right)
\end{equation}
\begin{equation}
J_z=\left(
\begin{array}{cccc}
 \frac{3}{2} & 0 & 0 & 0 \\
 0 & \frac{1}{2} & 0 & 0 \\
 0 & 0 & -\frac{1}{2} & 0 \\
 0 & 0 & 0 & -\frac{3}{2} \\
\end{array}
\right).
\end{equation}

The five Gamma matrices are defined as \cite{PhysRevB.69.235206}
\begin{equation}
\left\{
\begin{array}{l}
\Gamma^1=\frac{1}{\sqrt{3}} (J_y J_z+J_z J_y)\\
\Gamma^2=\frac{1}{\sqrt{3}} (J_z J_x+J_x J_z)\\
\Gamma^3=\frac{1}{\sqrt{3}} (J_x J_y+J_y J_x)\\
\Gamma^4=\frac{1}{\sqrt{3}} (J_x^2-J_y^2)\\
\Gamma^5=\frac{1}{3} (2 J_z^2-J_x^2-J_y^2)\\
\end{array}
\right..
\end{equation}
Clearly, $\{\Gamma^a,\Gamma^b\}=2\delta_{ab}\Gamma^0$ where $\Gamma^0$ is the 4 by 4 identity matrix.

Time reversal matrix $\gamma$ is defined as
\begin{equation}
\hat{\theta}c^{\dagger}_{\mathbf{k},\alpha}\hat{\theta}^{-1}=\sum_{\alpha'}c^{\dagger}_{-\mathbf{k},\alpha'}\gamma_{\alpha'\alpha},
\end{equation}
where $\hat{\theta}$ is the time-reversal operator.
The convention of time reversal matrix chosen in this work is
\cite{savary2017superconductivity}
\begin{equation}
\gamma=-i \Gamma^{13}=-\Gamma^1\Gamma^3,
\end{equation}
where $\Gamma^{ab}=\frac{1}{2i} [\Gamma^a,\Gamma^b]$ and $[...]$ is the anti-commutator.\cite{PhysRevB.69.235206}

The spin tensor\cite{PhysRevLett.117.075301,savary2017superconductivity} $M^{S m_S}$ is defined to satisfy the same rotation rule as angular momentum eigenstate $|S,m_S\rangle$.
Explicitly, if
$$
e^{-i \hat{\mathbf{S}}\cdot \mathbf{n} \theta}|S,m_S\rangle= \sum_{m_S'}R^{S}_{m_S' m_S}(\mathbf{n},\theta)|S,m_S'\rangle,
$$
$M^{S m_S}$ is defined to satisfy
$$
e^{-i \mathbf{J}\cdot \mathbf{n}\theta}M^{S m_S}e^{i \mathbf{J}\cdot\mathbf{n}\theta}=\sum_{m_S'}R^{S}_{m_S' m_S}(\mathbf{n},\theta)M^{S m_S'}
$$
for any three dimensional(3d) unit vector $\mathbf{n}$ and any angle $\theta$,
where $\mathbf{J}=(J_x,J_y,J_z)$ are the angular momentum matrices on the bases of the spin tensors.

Since the spin tensor is a rank-2 tensor, it can be viewed as the addition of two copies of spin basis.
In the spin-$\frac{3}{2}$ case, there are $1+3+5+7=16$ spin tensors with $S$ ranging from $0$ to $3$ and $m_S$ ranging from $-S$ to $S$.
The chosen expressions in this work are shown as the following\cite{PhysRevLett.117.075301,savary2017superconductivity}:
\begin{equation}
M^{00}=\Gamma^0
\end{equation}
\begin{equation}
\begin{array}{l}
M^{11}=\sqrt{\frac{2}{5}}(J_x+i J_y)\\
M^{10}=\sqrt{\frac{2}{5}}(-\sqrt{2} J_z)
\end{array}
\end{equation}
\begin{equation}
\begin{array}{l}
M^{22}=\frac{1}{\sqrt{2}}(-\Gamma^{4}-i \Gamma^{3})\\
M^{21}=\frac{1}{\sqrt{2}}(\Gamma^{2}+i \Gamma^{1})\\
M^{20}=- \Gamma^{5}
\end{array}
\end{equation}
\begin{equation}
\begin{array}{l}
M^{33}=\frac{1}{2} (-i \Gamma^{13}-\Gamma^{14}-\Gamma^{23}+i \Gamma^{24})\\
M^{32}=\frac{1}{\sqrt{2}}(-\Gamma^{35}+i \Gamma^{45})\\
M^{31}=\frac{\sqrt{3}}{2 \sqrt{5}} \left(-i \Gamma^{13}-\Gamma^{14}+\frac{2}{\sqrt{3}}\Gamma^{15}+\Gamma^{23}-i \Gamma^{24}-\frac{2 i}{\sqrt{3}}\Gamma^{25}\right)\\
M^{30}=\frac{1}{\sqrt{5}} (2\Gamma^{12}-\Gamma^{34})\\
\end{array},
\end{equation}
and $M^{S,-m_S}=(-1)^{m_S} (M^{S m_S})^{\dagger}$.
The spin tensors satisfy the orthogonal condition $\text{Tr}[(M^{S m_S})^{\dagger}M^{S' m_S'}]=4\delta_{S S'}\delta_{m_S m_S'}$.

Furthermore, $M^{S m_S}$ matrices satisfy the relation
\begin{equation}\label{Eqn:Mrelation2}
\delta_{s_1s_4}\delta_{s_2s_3}=\frac{1}{4}\sum_{S=0}^{3}\sum_{m_S=-S}^{S}(M^{S m_S}\gamma)_{s_1 s_2}(M^{S m_S}\gamma)^{\dagger}_{s_3s_4}
\end{equation}
with $s_{1,2,3,4}=\pm\frac{1}{2},\pm\frac{3}{2}$.

\section{Linearized gap equation and singlet-quintet mixing in Luttinger model}

\subsection{Green Functions of Luttinger model}
The Luttinger model shown in the main text can be rewritten as
$$
h(\mathbf{k})=
\xi_{\mathbf{k}}\Gamma^0+c_1\tilde{\mathbf{g}}_{\mathbf{k}}\cdot\mathbf{\Gamma},
$$
where $c_1 \tilde{\mathbf{g}}=\{c_1 g_{\mathbf{k},1},c_1 g_{\mathbf{k},2},c_1 g_{\mathbf{k},3},c_2 g_{\mathbf{k},4},c_2 g_{\mathbf{k},5}\}$, $\mathbf{\Gamma}=(\Gamma^1,\Gamma^2,...,\Gamma^5)$ and $\xi_{\mathbf{k}}= -\mu+\frac{1}{2m}k^2$.
That gives
\begin{equation}
\xi_{\mathbf{k}}=\xi_{-\mathbf{k}}\ ,\ \tilde{\mathbf{g}}_{\mathbf{k}}=\tilde{\mathbf{g}}_{-\mathbf{k}}
\end{equation}
and eigenenergies of $h(\mathbf{k})$ are
\begin{equation}
\xi_{\pm}(\mathbf{k})=\xi_{\mathbf{k}}\pm |c_1 \tilde{\mathbf{g}}_{\mathbf{k}}|.
\end{equation}

The Green functions of the Luttinger model are given by
\begin{equation}
\label{Ge_expression}
G_e(\mathbf{k},i \omega_n)=[i\omega_n- h(\mathbf{k})]^{-1}=[(i\omega_n-\xi_{\mathbf{k}})\Gamma^0-c_1\tilde{\mathbf{g}}_{\mathbf{k}}\cdot\mathbf{\Gamma}]^{-1}=\frac{(i\omega_n-\xi_{\mathbf{k}})\Gamma^0+c_1\tilde{\mathbf{g}}_{\mathbf{k}}\cdot\mathbf{\Gamma}}{(i\omega_n-\xi_{\mathbf{k}})^2-c_1^2|\tilde{\mathbf{g}}_{\mathbf{k}}|^2}
\end{equation}
and
\begin{equation}
\label{Gh_expression}
\gamma G_h(\mathbf{k},i \omega_n) \gamma^{-1}=\gamma[i\omega_n+ h^*(-\mathbf{k})]^{-1}\gamma^{-1}
=[i\omega_n+ h(\mathbf{k})]^{-1}
=\frac{(i\omega_n+\xi_{\mathbf{k}})\Gamma^0-c_1\tilde{\mathbf{g}}_{\mathbf{k}}\cdot\mathbf{\Gamma}}{(i\omega_n+\xi_{\mathbf{k}})^2-c_1^2|\tilde{\mathbf{g}}_{\mathbf{k}}|^2}
\end{equation}
for electrons and holes, respectively.
Here we use the fact that $h(\mathbf{k})$ is time-reversal invariant.

The Green functions can also be expressed in terms of projection operators $P_{\pm}(\mathbf{k})$,
 defined as
$$
P_{\pm}(\mathbf{k})\equiv \sum_{i=1}^{2}\left|\xi_{\pm}(\mathbf{k}),i\right\rangle\left\langle \xi_{\pm}(\mathbf{k}),i \right|
$$
in the subspace of the $\xi_{\pm}(\mathbf{k})$ bands, where $i$ stands for the double degeneracy of each band.
In the chosen bases, the matrix forms of $P_{\pm}(\mathbf{k})$ are
$$
P_{\pm}(\mathbf{k})=\frac{1}{2}\Gamma^0\pm \frac{c_1 \tilde{\mathbf{g}}_{\mathbf{k}}\cdot \mathbf{\Gamma}}{2 |c_1|\tilde{g}_{\mathbf{k}}}
$$
with $\tilde{g}_{\mathbf{k}}=|\tilde{\mathbf{g}}_{\mathbf{k}}|$.
Correspondingly,
$$
h(\mathbf{k})=\xi_+(\mathbf{k})P_+(\mathbf{k})+\xi_-(\mathbf{k})P_-(\mathbf{k}),
$$
and
\begin{equation}
\label{Ge_expression_projection}
G_e(\mathbf{k},i \omega_n)
=
[i\omega_n- h(\mathbf{k})]^{-1}
=
\frac{1}{i\omega_n-\xi_{+}}P_+(\mathbf{k})+\frac{1}{i\omega_n-\xi_{-}}P_-(\mathbf{k})
\end{equation}
\begin{equation}
\label{Gh_expression_projection}
\gamma G_h(\mathbf{k},i \omega_n) \gamma^{-1}
=
[i\omega_n+ h(\mathbf{k})]^{-1}
=
\frac{1}{i\omega_n+\xi_{+}}P_+(\mathbf{k})+\frac{1}{i\omega_n+\xi_{-}}P_-(\mathbf{k}),
\end{equation}
where $\xi_{\pm}=\xi_{\mathbf{k}}\pm |c_1|\tilde{g}_{\mathbf{k}}$.

The isotropic case corresponds $c_1=c_2$ in the above expressions.
Since $\hat{\mathbf{k}}\cdot\mathbf{J}$ commutes with $h(\mathbf{k})$ for $c_1=c_2$, energy eigenstates can be labeled with eigenvalues of $\hat{\mathbf{k}}\cdot\mathbf{J}$. In this case, the $\xi_+$ bands are $\frac{3}{2}$ bands if $c_1>0$, and $\frac{1}{2}$ bands if $c_1<0$.

\subsection{Expansion of interaction and gap function into different Channels}
This part follows Ref.\cite{savary2017superconductivity}.
Consider a three dimensional density-density interaction
\begin{equation}
\label{interaction_form_real_space}
H_{int}=\frac{1}{2}\int d^3 x \int d^3 x' \sum_{s_1,s_2,s_3,s_4}
U(\mathbf{x}-\mathbf{x}')\delta_{s_1s_4}\delta_{s_2s_3}c^{\dagger}_{\mathbf{x},s_1}c^{\dagger}_{\mathbf{x}',s_2} c_{\mathbf{x}',s_3}c_{\mathbf{x},s_4},
\end{equation}
where $s_{1,...,4}=\pm 1/2, \pm 2/3$.

After performing the Fourier transformation, we obtain
\begin{equation}
H_{int}=\frac{1}{2}\sum_{\mathbf{k}_3,\mathbf{k}_4,\mathbf{q}} \sum_{s_1,s_2,s_3,s_4}
V(\mathbf{q})\delta_{s_1s_4}\delta_{s_2s_3}c^{\dagger}_{\mathbf{k}_4-\mathbf{q},s_1}c^{\dagger}_{\mathbf{k}_3+\mathbf{q},s_2} c_{\mathbf{k}_3,s_3}c_{\mathbf{k}_4,s_4},
\end{equation}
where
$
c^{\dagger}_{\mathbf{x},s}=\frac{1}{\sqrt{\mathcal{V}}}\sum_{\mathbf{k}}c^{\dagger}_{\mathbf{k},s} e^{-i \mathbf{k}\cdot \mathbf{x}}
$
and
\begin{equation}
V(\mathbf{q})=\frac{1}{\mathcal{V}}\int d^3 r U(\mathbf{r}) e^{i \mathbf{q}\cdot \mathbf{r}}
\end{equation}
with the total volume $\mathcal{V}$.

Since Cooper pairs of superconductivity occurs for two electrons with opposite momenta, we only keep the terms
with $\mathbf{k_4}=-\mathbf{k_3}$ in the above interaction.
As a result, we can define $\mathbf{k}'=\mathbf{k_4}=-\mathbf{k_3}$ and $\mathbf{k}=\mathbf{k}'-\mathbf{q}$, which lead to
\begin{equation}
\label{interaction_form_momentum_space_compact}
H_{int}=\frac{1}{2}\sum_{\mathbf{k},\mathbf{k}'}\sum_{s_1,s_2,s_3,s_4}
V(\mathbf{k}-\mathbf{k}')\delta_{s_1s_4}\delta_{s_2s_3}c^{\dagger}_{\mathbf{k},s_1}c^{\dagger}_{-\mathbf{k},s_2} c_{-\mathbf{k}',s_3}c_{\mathbf{k}',s_4}.
\end{equation}
We generally denote $V(\mathbf{k}-\mathbf{k}')$ as $V(\mathbf{k},\mathbf{k}')$ and impose the $O(3)$ symmetry
on the interaction, $V(R\mathbf{k},R\mathbf{k}')=V(\mathbf{k},\mathbf{k}')$ for any $R\in O(3)$.
In addition, the Hermitian condition of interaction requires $V(\mathbf{k},\mathbf{k}')=V^*(\mathbf{k}',\mathbf{k})$.

Due to the $O(3)$ symmetry, $V(\mathbf{k},\mathbf{k}')$ can be expanded as
\begin{equation}
\label{expan_V_Y}
V(\mathbf{k},\mathbf{k}')=\sum_{l=0}^{\infty}\tilde{V}_l(|\mathbf{k}|,|\mathbf{k}'|)\frac{1}{2l+1}\sum_{m=-l}^{l}Y_{lm}(\hat{\mathbf{k}})Y^*_{lm}(\hat{\mathbf{k}}')
\end{equation}
with
\begin{equation}
\tilde{V}_l(|\mathbf{k}|,|\mathbf{k}'|)=\frac{1}{(4\pi)^2}\int d\hat{\mathbf{k}}\int d\hat{\mathbf{k}}'\sum_{m=-l}^{l}Y^*_{lm}(\hat{\mathbf{k}})Y_{lm}(\hat{\mathbf{k}}')V(\mathbf{k},\mathbf{k}').
\end{equation}
Here the spherical harmonic functions satisfy the orthogonal condition $\frac{1}{4\pi}\int d\hat{\mathbf{k}}Y^*_{lm}(\hat{\mathbf{k}})Y_{l'm'}(\hat{\mathbf{k}})=\delta_{ll'}\delta_{mm'}$.

With the relation (\ref{Eqn:Mrelation2}) and (\ref{expan_V_Y}),
\begin{equation}
H_{int}=\frac{1}{2}\sum_{\mathbf{k},\mathbf{k}'}\sum_{s_1,s_2,s_3,s_4}
\sum_{S=0}^{3}\sum_{m_S=-S}^{S}\sum_{l=0}^{\infty}\tilde{V}_l(|\mathbf{k}|,|\mathbf{k}'|)\frac{1}{2l+1}\sum_{m=-l}^{l}c^{\dagger}_{\mathbf{k},s_1}Y_{lm}(\hat{\mathbf{k}})(\frac{1}{2}M^{S m_S}\gamma)_{s_1 s_2}c^{\dagger}_{-\mathbf{k},s_2} c_{-\mathbf{k}',s_3}Y^*_{lm}(\hat{\mathbf{k}}')(\frac{1}{2}M^{S m_S}\gamma)^{\dagger}_{s_3s_4}c_{\mathbf{k}',s_4}.
\end{equation}

Since both $Y_{lm}(\hat{\mathbf{k}})$ and $(\frac{1}{2}M^{S m_S}\gamma)$ form irreducible representations (irreps) of $SO(3)$ group,
their product can be decomposed into new irreps with Clebsch–Gordan(C-G) coefficients as
\begin{equation}
Y_{lm}(\hat{\mathbf{k}})\frac{1}{2}M^{S m_S}
=\sum_{j=|l-S|}^{|l+S|}\sum_{m_j=-j}^{j}\left\langle l,S;j,m_j\right|\left.l,S;m,m_S\right\rangle N^{lS}_{jm_j}(\hat{\mathbf{k}}),
\end{equation}
where $\int \frac{d\Omega}{4\pi}\text{Tr}\{[N^{l'S'}_{j'm_j'}(\hat{\mathbf{k}})]^{\dagger}N^{lS}_{jm_j}(\hat{\mathbf{k}})\}=\delta_{ll'}\delta_{SS'}\delta_{jj'}\delta_{m_j m_j'}$ can be easily derived from the orthogonal conditions of $M$'s and $Y$'s.

With the above expansion, we have
$$
\sum_{m_S=-S}^{S}\sum_{m=-l}^{l}Y_{lm}(\hat{\mathbf{k}})(\frac{1}{2}M^{S m_S}\gamma)_{s_1 s_2} Y^*_{lm}(\hat{\mathbf{k}}')(\frac{1}{2}M^{S m_S}\gamma)^{\dagger}_{s_3s_4}
=
\sum_{m_S=-S}^{S}\sum_{m=-l}^{l}\sum_{j=|l-S|}^{|l+S|}\sum_{m_j=-j}^{j}\sum_{j'=|l-S|}^{|l+S|}\sum_{m_j'=-j}^{j}\left\langle l,S;j,m_j\right|\left.l,S;m,m_S\right\rangle
$$
\begin{equation}
\left\langle l,S;j',m_j'\right|\left.l,S;m,m_S\right\rangle^*
[N^{lS}_{jm_j}(\hat{\mathbf{k}})\gamma]_{s_1 s_2} [N^{lS}_{j'm_j'}(\hat{\mathbf{k}}')\gamma]^{\dagger}_{s_3s_4}
=
\sum_{j=|l-S|}^{|l+S|}\sum_{m_j=-j}^{j}[N^{lS}_{jm_j}(\hat{\mathbf{k}})\gamma]_{s_1 s_2} [N^{lS}_{jm_j}(\hat{\mathbf{k}}')\gamma]^{\dagger}_{s_3s_4},
\end{equation}
which gives rise to
\begin{equation}
H_{int}=\frac{1}{2}\sum_{\mathbf{k},\mathbf{k}'}\sum_{s_1,s_2,s_3,s_4}
\sum_{S=0}^{3}\sum_{l=0}^{\infty}\tilde{V}_l(|\mathbf{k}|,|\mathbf{k}'|)\frac{1}{2l+1}\sum_{j=|l-S|}^{|l+S|}\sum_{m_j=-j}^{j}c^{\dagger}_{\mathbf{k},s_1}[N^{lS}_{jm_j}(\hat{\mathbf{k}})\gamma]_{s_1 s_2}c^{\dagger}_{-\mathbf{k},s_2} c_{-\mathbf{k}',s_3}[N^{lS}_{jm_j}(\hat{\mathbf{k}}')\gamma]^{\dagger}_{s_3s_4}c_{\mathbf{k}',s_4}.
\end{equation}

Due to the anti-commutation relation of fermion operators, we have
$$
\sum_{\hat{\mathbf{k}}}\sum_{s_1 s_2}c^{\dagger}_{\mathbf{k},s_1}[N^{lS}_{jm_j}(\hat{\mathbf{k}})\gamma]_{s_1 s_2}c^{\dagger}_{-\mathbf{k},s_2}
=
-\sum_{\hat{\mathbf{k}}}\sum_{s_1 s_2}c^{\dagger}_{-\mathbf{k},s_2}[N^{lS}_{jm_j}(\hat{\mathbf{k}})\gamma]_{s_1 s_2}c^{\dagger}_{\mathbf{k},s_1}
=
\sum_{\hat{\mathbf{k}}}\sum_{s_1 s_2}c^{\dagger}_{\mathbf{k},s_1}[-N^{lS}_{jm_j}(-\hat{\mathbf{k}})\gamma]_{s_2 s_1}c^{\dagger}_{-\mathbf{k},s_2}
$$
\begin{equation}
\Leftrightarrow [-N^{lS}_{jm_j}(-\hat{\mathbf{k}})\gamma]^T=N^{lS}_{jm_j}(\hat{\mathbf{k}})\gamma ,
\end{equation}
which gives a constraint on the form of $N^{lS}_{jm_j}(\hat{\mathbf{k}})$.
Since $[N^{lS}_{jm_j}(-\hat{\mathbf{k}})\gamma]^T=(-1)^{l+S+1}N^{lS}_{jm_j}(\hat{\mathbf{k}})\gamma$,
it requires $l+S$ to be an even number. As a summary, the form of interaction term is given by
\begin{equation}
\label{interaction_form_3/2}
H_{int}=\frac{1}{2}\sum_{\mathbf{k},\mathbf{k}'}\sum_{s_1,s_2,s_3,s_4}
\sum_{S,l}'\tilde{V}_l(|\mathbf{k}|,|\mathbf{k}'|)\frac{1}{2l+1}\sum_{j=|l-S|}^{|l+S|}\sum_{m_j=-j}^{j}c^{\dagger}_{\mathbf{k},s_1}[N^{lS}_{jm_j}(\hat{\mathbf{k}})\gamma]_{s_1 s_2}c^{\dagger}_{-\mathbf{k},s_2} c_{-\mathbf{k}',s_3}[N^{lS}_{jm_j}(\hat{\mathbf{k}}')\gamma]^{\dagger}_{s_3s_4}c_{\mathbf{k}',s_4},
\end{equation}
where $\sum_{S,l}'$ is a part of $\sum_{S=0}^{3}\sum_{l=0}^{\infty}$ with $l+S$ being even.
We re-define $\tilde{V}_{lSj}(|\mathbf{k}|,|\mathbf{k}'|) = \tilde{V}_l(|\mathbf{k}|,|\mathbf{k}'|)/(2l+1)$
for the interaction in the $(l,S,j)$ channel.

The gap function $\Delta_{s_1 s_2}(\mathbf{k})$ is a $4\times 4$ matrix and can also be expanded as
\begin{equation}
\Delta_{s_1 s_2}(\mathbf{k})=\sum_{S=0}^{3}\sum_{m_S=-S}^{S}\sum_{l=0}^{\infty}\sum_{m=-l}^{l}\Delta^{lS}_{m m_S}(k)Y_{lm}(\hat{\mathbf{k}})(\frac{1}{2}M^{S m_S}\gamma)_{s_1 s_2}
\end{equation}
with the spherical harmonics and spin tensors.
Using C-G coefficients, we have
\begin{equation}
\Delta_{s_1 s_2}(\mathbf{k})=\sum_{S=0}^{3}\sum_{l=0}^{\infty}
\sum_{j=|l-S|}^{|l+S|}\sum_{m_j=-j}^{j} \Delta^{lS}_{j m_j}(k)[N^{lS}_{jm_j}(\hat{\mathbf{k}})\gamma]_{s_1 s_2},
\end{equation}
where $\Delta^{lS}_{j m_j}(k)=\sum_{m_S=-S}^{S}\sum_{m=-l}^{l}\left\langle l,S;j,m_j\right|\left.l,S;m,m_S\right\rangle \Delta^{lS}_{m m_S}(k)$.
Similarly, due to the anti-commutation relation of fermion operators,
only even $l+S$ terms are left, giving rise to
\begin{equation}
\Delta_{s_1 s_2}(\mathbf{k})=\sum_{S,l}'
\sum_{j=|l-S|}^{|l+S|}\sum_{m_j=-j}^{j} \Delta^{lS}_{j m_j}(k)[N^{lS}_{jm_j}(\hat{\mathbf{k}})\gamma]_{s_1 s_2}.
\end{equation}

\subsection{Derivation of Linearized Gap Equation}
In this part, we will derive the linearized gap equation. Consider a Hamiltonian with the form
\begin{equation}
H=\sum_{\mathbf{k},\alpha,\beta}c^{\dagger}_{\mathbf{k},\alpha}h_{\alpha\beta}(\mathbf{k})c_{\mathbf{k},\beta}+
\frac{1}{2}\sum_{\mathbf{k},\mathbf{k}',\alpha,\beta,\gamma,\delta}V_{\alpha\beta\gamma\delta}(\mathbf{k},\mathbf{k}')c^{\dagger}_{\mathbf{k},\alpha}c^{\dagger}_{-\mathbf{k},\beta}c_{-\mathbf{k}',\gamma}c_{\mathbf{k}',\delta}\ ,
\end{equation}
where the chemical potential is set to be the zero energy.

Define $b_{\alpha\beta}(\mathbf{k})=\langle c_{-\mathbf{k},\alpha} c_{\mathbf{k},\beta} \rangle$,
where $\langle A \rangle= \text{Tr}(e^{-\beta H} A)/\text{Tr}(e^{-\beta H})$ (
 The definition of average here is different from the average over angle in the main text).
The product of four fermionic operators can be simplified by neglecting the fluctuations around expectations (the mean-field approximation)
\begin{equation}
c^{\dagger}_{\mathbf{k},\alpha}c^{\dagger}_{-\mathbf{k},\beta}c_{-\mathbf{k}',\gamma}c_{\mathbf{k}',\delta}
\approx
b_{\beta\alpha}^*(\mathbf{k})c_{-\mathbf{k}',\gamma}c_{\mathbf{k}',\delta}
+
c^{\dagger}_{\mathbf{k},\alpha}c^{\dagger}_{-\mathbf{k},\beta} b_{\gamma\delta}(\mathbf{k}')
- b_{\beta\alpha}^*(\mathbf{k})b_{\gamma\delta}(\mathbf{k}'),
\end{equation}
where $b_{\beta\alpha}^*(\mathbf{k})= \langle c^{\dagger}_{\mathbf{k},\alpha}c^{\dagger}_{-\mathbf{k},\beta}\rangle$.
In the following discussion, mean-field approximation is always assumed.

The gap function $\Delta_{\alpha\beta}(\mathbf{k})$ is defined as
\begin{equation}
\label{General_Definition_Of_Delta}
\Delta_{\alpha\beta}(\mathbf{k})
=\sum_{\mathbf{k}'}\sum_{\gamma\delta}V_{\alpha\beta\gamma\delta}(\mathbf{k},\mathbf{k}')b_{\gamma\delta}(\mathbf{k}').
\end{equation}
The interaction Hamiltonian $H$ is expanded as
\begin{equation}
H_{int}
\approx
\frac{1}{2}
\left(
\sum_{\mathbf{k}'}\sum_{\gamma\delta} \Delta_{\delta\gamma}^*(\mathbf{k}')c_{-\mathbf{k}',\gamma}c_{\mathbf{k}',\delta}
+
\sum_{\mathbf{k}}\sum_{\alpha\beta}\Delta_{\alpha\beta}(\mathbf{k})c^{\dagger}_{\mathbf{k},\alpha}c^{\dagger}_{-\mathbf{k},\beta}
\right)
-
\frac{1}{2}\sum_{\mathbf{k},\mathbf{k}',\alpha,\beta,\gamma,\delta}V_{\alpha\beta\gamma\delta}(\mathbf{k},\mathbf{k}')b_{\beta\alpha}^*(\mathbf{k})b_{\gamma\delta}(\mathbf{k}')
\end{equation}
in the mean-field approximation, where $\Delta_{\delta\gamma}^*(\mathbf{k}')=\sum_{\mathbf{k}}\sum_{\alpha\beta}V_{\alpha\beta\gamma\delta}(\mathbf{k},\mathbf{k}')b_{\beta\alpha}^*(\mathbf{k})$. Here we have used the Hermitian condition of the interaction $V_{\alpha\beta\gamma\delta}^*(\mathbf{k},\mathbf{k}')=V_{\delta\gamma\beta\alpha}(\mathbf{k}',\mathbf{k})$.

With $\Psi_{\mathbf{k}}^{\dagger}=(c_{\mathbf{k}}^{\dagger},c_{-\mathbf{k}}^T)$, the Hamiltonian can be expressed in the BdG form
\begin{equation}
\label{General_BdG_H}
H\approx
\sum_{\mathbf{k}}^{\prime}
\Psi_{\mathbf{k}}^{\dagger}
\left(
\begin{matrix}
h(\mathbf{k}) & \Delta(\mathbf{k})\\
\Delta^{\dagger}(\mathbf{k}) & -h^T(-\mathbf{k})
\end{matrix}
\right)
\Psi_{\mathbf{k}}
-f
+\varepsilon_0' ,
\end{equation}
where $\sum_{\mathbf{k}}^{\prime}$ only covers half 1BZ(covering the whole 1BZ if counting its inversion partner),
$$
f=\frac{1}{2}\sum_{\mathbf{k},\mathbf{k}',\alpha,\beta,\gamma,\delta}V_{\alpha\beta\gamma\delta}(\mathbf{k},\mathbf{k}')b_{\beta\alpha}^*(\mathbf{k})b_{\gamma\delta}(\mathbf{k}') ,
$$
and
$
\varepsilon_0'=\sum_{\mathbf{k}}^{\prime}\text{Tr}[h(-\mathbf{k})]
$.

Plugging Eq.\ref{General_BdG_H} into definition of $b_{\gamma\delta}(\mathbf{k})$ and keeping $\Delta$ to first order on the right-hand side, we obtain
\begin{equation}
b_{\gamma\delta}(\mathbf{k})=\frac{1}{\beta}\sum_{\omega_n}\left[G_e(\mathbf{k},i\omega_n)\Delta(\mathbf{k})G_h(\mathbf{k},i\omega_n)\right]_{\delta\gamma}+O(\Delta^2).
\end{equation}
 Combining the above equation with Eq.\ref{General_Definition_Of_Delta}, the self-consistent linearized
 gap equation is derived as
\begin{equation}
\label{General_Linearized_Gap_Equation}
\Delta_{\alpha\beta}(\mathbf{k})= \frac{1}{\beta}\sum_{\mathbf{k}',\omega_n}\sum_{\gamma\delta}V_{\alpha\beta\gamma\delta}(\mathbf{k},\mathbf{k}')[G_e(\mathbf{k}',i \omega_n) \Delta(\mathbf{k}')G_h(\mathbf{k}',i \omega_n)]_{\delta\gamma} ,
\end{equation}
where $G_e(\mathbf{k},i \omega_n)=[i\omega_n-h(\mathbf{k})]^{-1}$ is the normal state Green function, $G_h(\mathbf{k},i \omega_n)=[i\omega_n+h^T(-\mathbf{k})]^{-1}$ and $\omega_n=(2n+1)\pi/\beta$ is the fermionic Matsubara frequency with $n$ being integers.
The superconducting transition temperature can be solved from Eq.(\ref{General_Linearized_Gap_Equation}).

\subsection{s-Wave Singlet and d-Wave Quintet Mixing in linearized gap equation}
In this section, we will show the singlet-quintet mixing is allowed in the above linearized gap equation for Luttinger Hamiltonian in the isotropic case and the symmetric SOC term $h_{SOC}$ defined in the main text plays a central role in this pairing mixing
mechanism.

If we choose the gap function on the left hand side of the gap equation to be s-wave singlet pairing, the gap equation will take the form
$$
\Delta^{00}_{00}(k)=\frac{1}{\beta}\int \frac{d\hat{\mathbf{k}}}{4\pi}\sum_{\alpha\beta}[N^{00}_{00}(\hat{\mathbf{k}})\gamma]^{\dagger}_{\beta \alpha}\sum_{\mathbf{k}',\omega_n}\sum_{\gamma\delta}V_{\alpha\beta\gamma\delta}(\mathbf{k},\mathbf{k}')[G_e(\mathbf{k}',i \omega_n) \Delta(\mathbf{k}')G_h(\mathbf{k}',i \omega_n)]_{\delta\gamma}.
$$
With the interaction form in Eq.(\ref{interaction_form_3/2}),
we have
\begin{equation}
\Delta^{00}_{00}(k)
=
\frac{1}{\beta}\sum_{\mathbf{k}',\omega_n}\tilde{V}_{000}(|\mathbf{k}|,|\mathbf{k}'|)\text{Tr}\{G_e(\mathbf{k}',i \omega_n) \Delta(\mathbf{k}')G_h(\mathbf{k}',i \omega_n)[N^{00}_{00}(\hat{\mathbf{k}'})\gamma]^{\dagger}\}
\end{equation}
\begin{equation}
\Rightarrow
\Delta^{00}_{00}(k)
=
\sum_{S,l}'
\sum_{j=|l-S|}^{|l+S|}\sum_{m_j=-j}^{j} \Delta^{lS}_{j m_j}(k)\frac{1}{\beta}\sum_{\mathbf{k}',\omega_n}\tilde{V}_{000}(|\mathbf{k}|,|\mathbf{k}'|)\text{Tr}\{G_e(\mathbf{k}',i \omega_n) [N^{lS}_{jm_j}(\hat{\mathbf{k}'})\gamma]G_h(\mathbf{k}',i \omega_n)[N^{00}_{00}(\hat{\mathbf{k}'})\gamma]^{\dagger}\}.
\end{equation}
The mixing between $\Delta^{00}_{00}(k)$ and $\Delta^{lS}_{j m_j}(k)$ in the above equation is
determined by
$$
\frac{1}{\beta}\sum_{\mathbf{k}',\omega_n}\tilde{V}_{000}(|\mathbf{k}|,|\mathbf{k}'|)\text{Tr}\{G_e(\mathbf{k}',i \omega_n) [N^{lS}_{jm_j}(\hat{\mathbf{k}'})\gamma]G_h(\mathbf{k}',i \omega_n)[N^{00}_{00}(\hat{\mathbf{k}'})\gamma]^{\dagger}\}.
$$

To simplify our discussion, we assume the $O(3)$ symmetry of non-interacting Hamiltonian. In this limit, we find
$$
\sum_{\hat{\mathbf{k}}',\omega_n}\text{Tr}\{G_e(\mathbf{k}',i \omega_n) [N^{lS}_{jm_j}(\hat{\mathbf{k}'})\gamma]G_h(\mathbf{k}',i \omega_n)[N^{00}_{00}(\hat{\mathbf{k}'})\gamma]^{\dagger}\}=0
$$
for $j\neq 0$ or $m_j\neq 0$ or $l$ is not even. Therefore, only the isotropic d-wave quintet pairing $\Delta^{22}_{0 0}(k)$ can be mixed into $\Delta^{00}_{0 0}(k)$ under the $O(3)$ symmetry.

Similarly, one can show the gap equation for isotropic d-wave quintet gap function $\Delta^{22}_{0 0}(k)$ is
$$
\Delta^{22}_{00}(k)
=
\sum_{S,l}'
\sum_{j=|l-S|}^{|l+S|}\sum_{m_j=-j}^{j} \Delta^{lS}_{j m_j}(k)\frac{1}{\beta}\sum_{\mathbf{k}',\omega_n}\tilde{V}_{220}(|\mathbf{k}|,|\mathbf{k}'|)\text{Tr}\{G_e(\mathbf{k}',i \omega_n) [N^{lS}_{jm_j}(\hat{\mathbf{k}'})\gamma]G_h(\mathbf{k}',i \omega_n)[N^{22}_{00}(\hat{\mathbf{k}'})\gamma]^{\dagger}\},
$$
and only $\Delta^{00}_{00}(k)$ can be mixed into
$\Delta^{22}_{0 0}(k)$ under the $O(3)$ symmetry.
Thus, for the chosen $O(3)$ invariant interaction and a generic $O(3)$ invariant non-interacting Hamiltonian,
$(0,0,0)$ channel is only allowed to mix with $(2,2,0)$ channel and vice verse.

Next, we will show what terms of the non-interacting Hamiltonian that are essential for the existence of the mixing.
The general form of the $O(3)$ invariant non-interacting Hamiltonian reads
\begin{equation}
h(\mathbf{k})=2 f_1(k)N^{00}_{00}(\hat{\mathbf{k}})-2 f_2(k)N^{22}_{00}(\hat{\mathbf{k}}) ,
\end{equation}
where $N^{00}_{00}(\hat{\mathbf{k}})$ and $N^{22}_{00}(\hat{\mathbf{k}})$ are chosen to be Hermitian and $f_{1,2}(k)$ are arbitrary real functions of magnitude of $\mathbf{k}$.
In terms of $\Gamma$ matrices,
the general form of the $O(3)$ invariant non-interacting Hamiltonian reads
\begin{equation}
N^{00}_{00}(\hat{\mathbf{k}})=\frac{1}{2}\Gamma^0
\end{equation}
and
\begin{equation}
N^{22}_{00}(\hat{\mathbf{k}})
=\sum_{m,m_S=-2}^{2}\left\langle 2,2; m, m_S\right|\left. 2,2; 0,0\right\rangle Y_{2m}(\hat{\mathbf{k}})\frac{1}{2}M^{2 m_S}
=
\sum_{m,m_S=-2}^{2}\delta_{m,-m_S}\frac{(-1)^m}{\sqrt{5}} Y_{2m}(\hat{\mathbf{k}})\frac{1}{2}M^{2 m_S}
\end{equation}
\begin{equation}
\Rightarrow
N^{22}_{00}(\hat{\mathbf{k}})
=
\sum_{m=-2}^{2}\frac{(-1)^m}{\sqrt{5}} Y_{2m}(\hat{\mathbf{k}})\frac{1}{2}M^{2 ,-m}
=
-\frac{1}{2}\hat{\mathbf{g}}\cdot\mathbf{\Gamma} ,
\end{equation}
from which one can see that $N^{22}_{00}(\hat{\mathbf{k}})$ follows the form of symmetric spin-orbit coupling in the isotropic case. Therefore, we consider the Hamiltonian with the form
\begin{equation}
\label{isotropic_non_interacting_Hamiltonian}
h(\mathbf{k})=f_1(k)\Gamma^{0}+f_2(k)\hat{\mathbf{g}}\cdot\mathbf{\Gamma} ,
\end{equation}
which leads to the Green functions
\begin{equation}
G_{e}(i\omega,\mathbf{k})=(i\omega-h(\mathbf{k}))^{-1}=\frac{i\omega-f_1(k)+f_2(k)\hat{\mathbf{g}}\cdot\mathbf{\Gamma}}{(i\omega-f_1(k))^2-f_2^2(k)},
\end{equation}
and
\begin{equation}
\gamma G_{h}(i\omega,\mathbf{k})\gamma^{\dagger}= (i\omega+h(\mathbf{k}))^{-1}=\frac{i\omega+f_1(k)-f_2(k)\hat{\mathbf{g}}\cdot\mathbf{\Gamma}}{(i\omega+f_1(k))^2-f_2^2(k)}.
\end{equation}

With the above form of Green functions, we have
\begin{equation}
\gamma G_{h}(i\omega,\mathbf{k})\gamma^{\dagger}[N_{00}^{00}(\hat{\mathbf{k}})]^{\dagger}G_{e}(i\omega,\mathbf{k})
=
\frac{-(\omega^2+f_1(k)^2+f_2(k)^2)[N^{00}_{00}(\hat{\mathbf{k}})]^{\dagger}-2f_1(k)f_2(k)[N^{22}_{00}(\hat{\mathbf{k}})]^{\dagger}}{[(i\omega-f_1(k))^2-f_2^2(k)][(i\omega+f_1(k))^2-f_2^2(k)]}
\end{equation}
\begin{equation}
\gamma G_{h}(i\omega,\mathbf{k})\gamma^{\dagger}[N_{22}^{00}(\hat{\mathbf{k}})]^{\dagger}G_{e}(i\omega,\mathbf{k})
=
\frac{-(\omega^2+f_1(k)^2+f_2(k)^2)[N^{22}_{00}(\hat{\mathbf{k}})]^{\dagger}-2f_1(k)f_2(k)[N^{00}_{00}(\hat{\mathbf{k}})]^{\dagger}}{[(i\omega-f_1(k))^2-f_2^2(k)][(i\omega+f_1(k))^2-f_2^2(k)]}
\end{equation}
Plugging into the linearized gap equation and using the orthonormal condition for $N$'s,
\begin{equation}
\Delta^{00}_{00}(k)
=
\frac{1}{\beta}\sum_{\mathbf{k}',\omega_n}\tilde{V}_{000}(|\mathbf{k}|,|\mathbf{k}'|)\frac{-(\omega^2+f_1(k)^2+f_2(k)^2) \Delta^{00}_{00}(k)-2f_1(k)f_2(k) \Delta^{22}_{0 0}(k)}{[(i\omega-f_1(k))^2-f_2^2(k)][(i\omega+f_1(k))^2-f_2^2(k)]},
\end{equation}
\begin{equation}
\Delta^{22}_{00}(k)
=
\frac{1}{\beta}\sum_{\mathbf{k}',\omega_n}\tilde{V}_{220}(|\mathbf{k}|,|\mathbf{k}'|)\frac{-(\omega^2+f_1(k)^2+f_2(k)^2) \Delta^{22}_{00}(k)-2f_1(k)f_2(k) \Delta^{00}_{0 0}(k)}{[(i\omega-f_1(k))^2-f_2^2(k)][(i\omega+f_1(k))^2-f_2^2(k)]}.
\end{equation}

Therefore, non-trivial solutions of the above gap equations require (1) non-zero interaction parameters $\tilde{V}_{000},\tilde{V}_{220}$ and (2) non-zero $f_{1,2}(k)$.
The condition (2) suggests the essential role of symmetric SOC.

The above analysis can be carried out in a more compact form similar to the case of singlet-triplet mixing in non-centrosymmetric superconductors, as discussed in Ref.\cite{Frigeri2004}.
We will choose the isotropic Luttinger Hamiltonian ($c_1=c_2$) with the $O(3)$ invariant interaction Eq. (\ref{interaction_form_momentum_space_compact}) and choose the gap function as
\begin{equation}
\Delta(\mathbf{k})=\varphi(k)\Gamma^0\gamma+\mathbf{d}(\mathbf{k})\cdot\mathbf{\Gamma}\gamma
\end{equation} with only s-wave singlet pairing and a generic d-wave quintet pairing.
We omit the triplet and septet channels because they are parity-odd and the chosen Luttinger model is centrosymmetric.
In this case, the linearized gap equation reads
\begin{equation}
\Delta(\mathbf{k})\gamma^{\dagger}= k_B T\sum_{\mathbf{k}',\omega_n}V(\mathbf{k},\mathbf{k}')G_e(\mathbf{k}',i \omega_n) (\varphi(k')\Gamma^0 +\mathbf{d}(\mathbf{k}')\cdot\mathbf{\Gamma})\gamma G_h(\mathbf{k}',i \omega_n)\gamma^{\dagger},
\end{equation}
and lead to two coupled equations for $\varphi(k)$ and $\mathbf{d}(\mathbf{k})$
\begin{equation}
\varphi(k)=k_B T\sum_{\omega_n,\mathbf{k}'}\frac{V(\mathbf{k},\mathbf{k}')}{b(\mathbf{k}',i\omega_n)}[(-c_1^2 g_{\mathbf{k}'}^2-\xi_{\mathbf{k}'}^2-\omega_n^2)\varphi(k')+2c_1 \mathbf{d}(\mathbf{k}')\cdot \mathbf{g}_{\mathbf{k}'}\xi_{\mathbf{k}'}]
\end{equation}
\begin{equation}
\mathbf{d}(\mathbf{k})=k_B T\sum_{\omega_n,\mathbf{k}'}\frac{V(\mathbf{k},\mathbf{k}')}{b(\mathbf{k}',i\omega_n)}[(-c_1^2 g_{\mathbf{k}'}^2-\xi_{\mathbf{k}'}^2-\omega_n^2)\mathbf{d}(\mathbf{k}')+2c_1^2(g_{\mathbf{k}'}^2\mathbf{d}(\mathbf{k}')-\mathbf{g}_{\mathbf{k}'} \mathbf{d}(\mathbf{k}')\cdot\mathbf{g}_{\mathbf{k}'})+2c_1 \mathbf{g}_{\mathbf{k}'}\xi_{\mathbf{k}'}\varphi(k')],
\end{equation}
where $b(\mathbf{k},i\omega_n)=[(\xi_{\mathbf{k}}-i\omega_n)^2-g_{\mathbf{k}}^2 c_1^2][(\xi_{\mathbf{k}}+i\omega_n)^2-g_{\mathbf{k}}^2 c_1^2]$.
In the equation of $\varphi(k)$, since the s-wave pairing is isotropic $\varphi(k)=\int d\Omega\varphi(k)/(4\pi)$, the mixing term can be re-written as
\begin{equation}
\int \frac{d\Omega}{4\pi}\sum_{\omega_n,\mathbf{k}'}\frac{V(\mathbf{k},\mathbf{k}')}{b(\mathbf{k}',i\omega_n)}2c_1 \mathbf{d}(\mathbf{k}')\cdot \mathbf{g}_{\mathbf{k}'}\xi_{\mathbf{k}'}
=
\sum_{\mathbf{k}'}f_V(k,k')c_1 \mathbf{d}(\mathbf{k}')\cdot \mathbf{g}_{\mathbf{k}'},
\end{equation}
where $f_V(k,k')=\sum_{\omega_n}\int \frac{d\Omega}{4\pi} \frac{2V(\mathbf{k},\mathbf{k}')\xi_{\mathbf{k}'}}{b(\mathbf{k}',i\omega_n)}$ is a $O(3)$ invariant function.
From the above expression, it is clear that the mixing term will vanish for a zero symmetric SOC term $h_{SOC}$
($c_1=0$). In addition, we can see that the vector $\mathbf{d}(\mathbf{k})$ should contain a component parallel to the vector $\mathbf{g}_{\mathbf{k}}$ for a non-zero mixing term. The above derived coupled gap equations are quite similar to those for singlet-triplet mixing in non-centrosymmetric SCs\cite{Frigeri2004}.
Given the d-wave nature of $\mathbf{g}_{\mathbf{k}}$ in $h_{SOC}$, we conclude that
only d-wave component in the quintet channel can be mixed into s-wave singlet pairing.

The above analysis actually presents us a minimal model that can be chosen for this problem: the $O(3)$ invariant interaction only contains $(0,0,0)$ and $(2,2,0)$ channels with two parameters $V_0$ and $V_1$ discussed in the main text and $O(3)$ invariant non-interacting Hamiltonian with the form in Eq. (\ref{isotropic_non_interacting_Hamiltonian}).

\subsection{Justification of the interaction term}
In the main text, we present our linearized gap equation based on a simplified interaction form with
two parameters $V_0$ and $V_1$ in s-wave singlet and d-wave quintet channels. In this section, we will justify this form of interaction from a more realistic interaction. Here we consider a screened Coulomb-like potential, which has been used in Ref.\cite{savary2017superconductivity}. We notice that such interaction can be generated
by the electric polarization of the optical phonon modes and is used to explain the critical temperature of superconductivity in this superconducting material with the extremely low
density of carriers.\cite{savary2017superconductivity}

Assume $U(\mathbf{x}-\mathbf{x}')$ in Eq.\ref{interaction_form_real_space} has the form of an isotropic and inversion invariant screened Coulomb-like potential
\begin{equation}
U(\mathbf{x}-\mathbf{x}')=\frac{A}{4\pi}\frac{e^{- B |\mathbf{x}-\mathbf{x}'|}}{|\mathbf{x}-\mathbf{x}'|}
\end{equation}
with $B>0$ and $A<0$ (attractive interaction).

Its Fourier transformation has the form
\begin{equation}
V(\mathbf{q})=\frac{1}{\mathcal{V}}\int d^3 r U(\mathbf{r}) e^{i \mathbf{k}\cdot \mathbf{r}}=\frac{1}{\mathcal{V}}\frac{A}{|\mathbf{q}|^2+B^2},
\end{equation}
which can be used for
$V(\mathbf{k},\mathbf{k}')$ in Eq.\ref{interaction_form_momentum_space_compact}
($V(\mathbf{q}=\mathbf{k}-\mathbf{k}')$).

We are only interested in the form of the interaction in the $(0,0,0)$ and $(2,2,0)$ channels, and thus
expand the interaction as
\begin{equation}
H_{int}=\frac{1}{2}\sum_{\mathbf{k},\mathbf{k}'}
\left[
\tilde{V}_{000} c^{\dagger}_{\mathbf{k}}\left(\frac{1}{2}\Gamma_0\gamma\right)\left(c^{\dagger}_{-\mathbf{k}}\right)^T\left(c_{-\mathbf{k}'}\right)^T\left(\frac{1}{2}\Gamma_0\gamma\right)^{\dagger}c_{\mathbf{k}'}
+\tilde{V}_{220} c^{\dagger}_{\mathbf{k}}\left(-\frac{1}{2}\hat{\mathbf{g}}_{\mathbf{\mathbf{k}}}\cdot\mathbf{\Gamma}\gamma\right)\left(c^{\dagger}_{-\mathbf{k}}\right)^T\left(c_{-\mathbf{k}'}\right)^T\left(-\frac{1}{2}\hat{\mathbf{g}}_{\mathbf{\mathbf{k}}'}\cdot\mathbf{\Gamma}\gamma\right)^{\dagger}c_{\mathbf{k}'}
\right],
\end{equation}
where
\begin{equation}
\tilde{V}_{000}(|\mathbf{k}|,|\mathbf{k}'|)=\frac{1}{(4\pi)^2}\int d\hat{\mathbf{k}}\int d\hat{\mathbf{k}}'V(\mathbf{k}-\mathbf{k}')
=
\frac{A \ln \left(\frac{4 k k'}{B^2+(k-k')^2}+1\right)}{4 k k' \mathcal{V}}
\end{equation}
,
\begin{equation}
\tilde{V}_{220}(|\mathbf{k}|,|\mathbf{k}'|)=\frac{1}{(4\pi)^2}\int d\hat{\mathbf{k}}\int d\hat{\mathbf{k}}'\frac{5}{2}(-1+3 \hat{\mathbf{k}}\cdot \hat{\mathbf{k}}')V(\mathbf{k}-\mathbf{k}')
\end{equation}
\begin{equation}
=
\frac{5 A \left(2 k k' \left(-3 \left(B^2+k^2+k'^2\right)-k k' \ln \left(\frac{4 k k'}{B^2+(k-k')^2}+1\right)\right)+3 \left(B^2+k^2+k'^2\right)^2 \tanh ^{-1}\left(\frac{2 k k'}{B^2+k^2+k'^2}\right)\right)}{16 k^3 k'^3 \mathcal{V}}
\end{equation}
with
$k=|\mathbf{k}|$
and
$k'=|\mathbf{k}'|$.

Assuming $k/B\sim k'/B\ll 1$, we find
\begin{equation}
\tilde{V}_0(|\mathbf{k}|,|\mathbf{k}'|)\approx\frac{A}{B^2\mathcal{V}}
\end{equation}
and
\begin{equation}
\tilde{V}_2(|\mathbf{k}|,|\mathbf{k}'|)\approx
\frac{8 A}{3 B^6 \mathcal{V}}k'^2 k^2
\end{equation}
up to the leading order. In the above limit, we notice that $\tilde{V}_2\ll \tilde{V}_0$.

With $V_0=A/B^2$ and $V_1=8 A/ (3 B^6 a^4)$, the interaction term should take the form
\begin{equation}
\label{chosen_interaction_form}
H_{int}=\frac{1}{2\mathcal{V}}\sum_{\mathbf{k},\mathbf{k}'}
\left[
V_0 c^{\dagger}_{\mathbf{k}}\left(\frac{1}{2}\Gamma_0\gamma\right)\left(c^{\dagger}_{-\mathbf{k}}\right)^T\left(c_{-\mathbf{k}'}\right)^T\left(\frac{1}{2}\Gamma_0\gamma\right)^{\dagger}c_{\mathbf{k}'}
+V_1 c^{\dagger}_{\mathbf{k}}\left(\frac{a^2}{2}\mathbf{g}_{\mathbf{\mathbf{k}}}\cdot\mathbf{\Gamma}\gamma\right)\left(c^{\dagger}_{-\mathbf{k}}\right)^T\left(c_{-\mathbf{k}'}\right)^T\left(\frac{a^2}{2}\mathbf{g}_{\mathbf{\mathbf{k}}'}\cdot\mathbf{\Gamma}\gamma\right)^{\dagger}c_{\mathbf{k}'}
\right],
\end{equation}
which is the same as that used in the main text.
Since the values of $A$ and $B$ are material dependent, we just regard $V_0$ and $V_1$ as two independent parameters in the main text for simplicity.
Moreover, we assume that the energy cut-off for attractive $V_{0,1}$ to be $\epsilon_c$.
$V_{0,1}$ are always assumed to be attractive unless specified otherwise.

\subsection{Solutions of the coupled linearized gap equation}
According to the gap function Eq. (\ref{General_Definition_Of_Delta}) and the interaction form Eq. (\ref{chosen_interaction_form}),
we can write the specific gap functions
\begin{equation}
\label{paring_form}
\Delta_{\alpha\beta}(\mathbf{k})
=\sum_{\mathbf{k}'}\sum_{\gamma\delta}V_{\alpha\beta\gamma\delta}(\mathbf{k},\mathbf{k}')b_{\gamma\delta}(\mathbf{k}')
=\Delta_0(\frac{\Gamma^0}{2}\gamma)_{\alpha\beta}
+
\Delta_1( \frac{a^2\mathbf{g}_{\mathbf{k}}\cdot\mathbf{\Gamma}}{2}\gamma)_{\alpha\beta},
\end{equation}
where
\begin{equation}
\Delta_0=\frac{V_0}{\mathcal{V}} \sum_{\mathbf{k}'}\sum_{\gamma\delta} [\frac{\Gamma^0}{2}\gamma]^{\dagger}_{\gamma\delta}b_{\gamma\delta}(\mathbf{k}')
\end{equation}
and
\begin{equation}
\Delta_1=\frac{V_1}{\mathcal{V}} \sum_{\mathbf{k}'}\sum_{\gamma\delta} [ \frac{a^2\mathbf{g}_{\mathbf{k}'}\cdot\mathbf{\Gamma}}{2}\gamma]^{\dagger}_{\gamma\delta}b_{\gamma\delta}(\mathbf{k}').
\end{equation}

With the linearized gap equation (\ref{General_Linearized_Gap_Equation}), we find the coupled linearized gap equations in the singlet and qunitet channels take the form
\begin{align}
\Delta_0=\frac{V_0}{\beta\mathcal{V}}\sum_{\mathbf{k},\omega_n}\text{Tr}\left[G_e(\mathbf{k},i \omega_n) (\Delta_0\frac{\Gamma^0}{2}\gamma
+\Delta_1 \frac{a^2\mathbf{g}_{\mathbf{k}}\cdot\mathbf{\Gamma}}{2}\gamma)G_h(\mathbf{k},i \omega_n)(\frac{\Gamma^0}{2}\gamma)^{\dagger}\right]
\\
\Delta_1=\frac{V_1}{\beta\mathcal{V}}\sum_{\mathbf{k},\omega_n}\text{Tr}\left[G_e(\mathbf{k},i \omega_n) (\Delta_0\frac{\Gamma^0}{2}\gamma
+\Delta_1 \frac{a^2\mathbf{g}_{\mathbf{k}}\cdot\mathbf{\Gamma}}{2}\gamma)G_h(\mathbf{k},i \omega_n)( \frac{a^2\mathbf{g}_{\mathbf{k}}\cdot\mathbf{\Gamma}}{2}\gamma)^{\dagger}\right].
\end{align}

With the Green functions in Eq.\ref{Ge_expression} and Eq.\ref{Gh_expression}, the coupled gap equations are re-written as
$$
\left(
\begin{array}{c}
\Delta_{0}\\
\Delta_{1}
\end{array}
\right)
=
\sum_{\mathbf{k},\omega_n}\frac{1}{\beta \mathcal{V} b(\mathbf{k},i\omega_n)}\left(
\begin{array}{cc}
-V_0 (k^4 Q_c^2+\xi^2+\omega_n^2)& V_0 2 k^4 a^2 (c_1 Q_1^2+c_2 Q_2^2) \xi\\
V_1 2 k^4 a^2 (c_1 Q_1^2+c_2 Q_2^2) \xi & -V_1 k^4 a^4 [(k^4 Q_c^2+\xi^2+\omega_n^2)- 2k^4(Q_c^2 -(c_1 Q_1^2+c_2 Q_2^2)^2)]
\end{array}
\right)
\left(
\begin{array}{c}
\Delta_{0}\\
\Delta_{1}
\end{array}
\right),
$$
where $b(\mathbf{k},i\omega_n)=(\xi_-^2+\omega_n^2)(\xi_+^2+\omega_n^2)$ , $\xi_{\pm}=\frac{k^2}{2m_{\pm}}-\mu$, $m_{\pm}=m/(1\pm 2mQ_c)$
,$Q_c=\sqrt{c_1^2 Q_1^2+c_2^2 Q_2^2}$, $Q_1=\sqrt{\hat{g}^2_{1}+\hat{g}^2_{2}+\hat{g}^2_{3}}$, $Q_2=\sqrt{\hat{g}^2_{4}+\hat{g}^2_{5}}$ and $\hat{g}_i=g_i/k^2$.
It is easy to see that the mixing is zero if $c_1=c_2=0$, which means the symmetric SOC is essential.

The above equations would be easier to deal with if expressed in terms of the projection operators.
With Eq.\ref{Ge_expression_projection} and Eq.\ref{Gh_expression_projection}, we have
$$
\Delta_{0}=-\frac{V_0}{\beta\mathcal{V}}\left\{\sum_{\mathbf{k},\omega_n}\frac{1}{\omega_n^2+\xi_{+}^2}(\frac{\Delta_{0}}{2}+\frac{\Delta_{1} a^2 k^2 }{2}\frac{c_1 Q_1^2+c_2 Q_2^2}{Q_c})
+
\sum_{\mathbf{k},\omega_n}\frac{1}{\omega_n^2+\xi_{-}^2}(\frac{\Delta_{0}}{2}-\frac{\Delta_{1} a^2 k^2 }{2}\frac{c_1 Q_1^2+c_2 Q_2^2}{Q_c})
\right\}
$$
$$
\Delta_{1}=-\frac{V_1}{\beta\mathcal{V}}\left\{\sum_{\mathbf{k},\omega_n}\frac{1}{\omega_n^2+\xi_{+}^2}(\frac{\Delta_{0} a^2 k^2 }{2}\frac{c_1 Q_1^2+c_2 Q_2^2}{Q_c}+\frac{\Delta_{1} a^4 k^4 }{2}\left(\frac{c_1 Q_1^2+c_2 Q_2^2}{Q_c}\right)^2)
\right.
$$
$$
+
\sum_{\mathbf{k},\omega_n}\frac{1}{\omega_n^2+\xi_{-}^2}(\frac{-\Delta_{0} a^2 k^2 }{2}\frac{c_1 Q_1^2+c_2 Q_2^2}{Q_c}+\frac{\Delta_{1} a^4 k^4 }{2}\left(\frac{c_1 Q_1^2+c_2 Q_2^2}{Q_c}\right)^2)
\left.
-
\sum_{\mathbf{k},\omega_n}\frac{1}{(i\omega_n-\xi_+)(i\omega_n+\xi_-)}\frac{(c_1-c_2)^2Q_1^2Q_2^2k^4 a^4 \Delta_{1}}{Q_c^2}
\right\}.
$$

We consider the limit $\epsilon_c/2 Q_c k_F^2\ll 1$ and $|\epsilon_c/\mu|\ll 1$, where $\epsilon_c$ labels the energy range for the momentum summation around the chemical potential $\mu$ and is an energy scale much smaller than SOC strength and chemical potential.
In the continuous limit, the momentum summation can thus be written as
$$
\frac{1}{\mathcal{V}}\sum_{\mathbf{k}}=\int \frac{d\Omega}{4\pi}N_{-}(0)\int_{-\epsilon_c}^{\epsilon_c}d\xi_{-}\sqrt{1+\frac{\xi_{-}}{\mu}}+\int \frac{d\Omega}{4\pi}N_{+}(0)\int_{-\epsilon_c}^{\epsilon_c}d\xi_{+}\sqrt{1+\frac{\xi_{+}}{\mu}},
$$
where
$
N_{\pm}(0)\equiv (2\pi)^{-3} 4\pi |m_{\pm}| \text{Re}[\sqrt{2m_{\pm} \mu}]
$ is the density of states for $\xi_\pm$ bands at Fermi energy without spin degeneracy.

Given
$\epsilon_c/(2 Q_c k_F^2)\ll 1$, the following four expressions
$$
\int_{-\epsilon_c}^{\epsilon_c}d\xi_{-}\sum_{\omega_n}\frac{1}{\beta(\xi_+^2+\omega_n^2)}F(\mathbf{k})
=
O[\frac{\epsilon_c}{2k_{F,-}^2Q_c}]\int_{-1}^{1}d(\frac{\xi_-}{\epsilon_c})\frac{F(\mathbf{k})}{2},
$$
$$
\int_{-\epsilon_c}^{\epsilon_c}d\xi_{+}\sum_{\omega_n}\frac{1}{\beta(\xi_-^2+\omega_n^2)}F(\mathbf{k})
=
O[\frac{\epsilon_c}{2k_{F,+}^2Q_c}]\int_{-1}^{1}d(\frac{\xi_+}{\epsilon_c})\frac{F(\mathbf{k})}{2},
$$
$$
\int_{-\epsilon_c}^{\epsilon_c}d\xi_{-}\sum_{\omega_n}\frac{1}{(i\omega_n-\xi_+)(i\omega_n+\xi_-)}F(\mathbf{k})
=
O[\frac{\epsilon_c}{2k_{F,-}^2Q_c}]\int_{-1}^{1}d(\frac{\xi_-}{\epsilon_c})F(\mathbf{k}),
$$
$$
\int_{-\epsilon_c}^{\epsilon_c}d\xi_{+}\sum_{\omega_n}\frac{1}{(i\omega_n-\xi_+)(i\omega_n+\xi_-)}F(\mathbf{k})
=
O[\frac{\epsilon_c}{2k_{F,+}^2Q_c}]\int_{-1}^{1}d(\frac{\xi_+}{\epsilon_c})F(\mathbf{k})
$$
are of order $\frac{\epsilon_c}{2k_{F,-}^2Q_c}$, and thus can be dropped.
With the above approximations as well as low transition temperature assumption $1/(\beta\epsilon_c)\ll 1$, the coupled linearized gap equation can be simplified as
\begin{equation}
\label{linearized_gap_eqn_approx}
\left(
\begin{array}{c}
\tilde{\Delta}_{0}\\
\tilde{\Delta}_{1}
\end{array}
\right)
=
x\left(
\begin{array}{cc}
\frac{1}{2}\lambda_{0}y_1 & \frac{1}{2}\lambda_{0} y_2\\
\frac{1}{2}\tilde{\lambda}_{1} y_2& \frac{1}{2}\tilde{\lambda}_{1} y_3
\end{array}
\right)
\left(
\begin{array}{c}
\tilde{\Delta}_{0}\\
\tilde{\Delta}_{1}
\end{array}
\right)
\end{equation}
up to the leading order of
$\epsilon_c/(2 Q_c k_F^2)$, $|\epsilon_c/\mu|$ and $1/(\beta\epsilon_c)$.
Here $y_{1,2,3}$ are given by
\begin{equation}
\label{eq_y1}
y_1\equiv \int \frac{d\Omega}{4\pi}\text{Re}\left[\frac{1}{(1-2m Q_c)^{3/2}}+\frac{1}{(1+2mQ_c)^{3/2}}\right]\geq 0,
\end{equation}
\begin{equation}
\label{eq_y2}
y_2\equiv \int \frac{d\Omega}{4\pi}\text{Re}\left[-\frac{1}{(1-2m Q_c)^{5/2}}+\frac{1}{(1+2mQ_c)^{5/2}}\right]\frac{|c_1|Q_1^2 + |c_2|Q_2^2}{ Q_c},
\end{equation}
\begin{equation}
\label{eq_y3}
y_3\equiv \int \frac{d\Omega}{4\pi}\text{Re}\left[\frac{1}{(1-2m Q_c)^{7/2}}+\frac{1}{(1+2mQ_c)^{7/2}}\right]\left(\frac{Q_1^2  |c_1|+Q_2^2  |c_2|}{ Q_c}\right)^2 \geq 0,
\end{equation}
$x=\ln(2e^{\bar{\gamma}}\beta \epsilon_c/\pi)$ with $\bar{\gamma}$ being the Euler constant
,
$\lambda_i\equiv -V_i N_0>0$ with $N_0=(2\pi)^{-3} 4\pi |m| \sqrt{2m \mu}$
,
$\tilde{\Delta}_{0}\equiv\Delta_{0}\text{sgn}(c_1)$
, $\tilde{\Delta}_{1}\equiv\Delta_{1}(2m \mu a^2)$
, $\tilde{\lambda}_1\equiv\lambda_1(2m \mu a^2)^2$
 and $c_1c_2>0, m<0,\mu<0$.

The coupled Eqs. (\ref{linearized_gap_eqn_approx}) can be solved as an eigen problem and
the corresponding eigen-values are
\begin{equation}
\frac{1}{x_1}=\frac{1}{4} \left(-\sqrt{(\lambda_0 y_{1}-\tilde{\lambda}_1 y_3)^2+4 \lambda_0 \tilde{\lambda}_1 y_{2}^2}+\lambda_0 y_{1}+\tilde{\lambda}_1 y_{3}\right)
\end{equation}
and
\begin{equation}
\frac{1}{x_2}=\frac{1}{4} \left(\sqrt{(\lambda_0 y_{1}-\tilde{\lambda}_1 y_3)^2+4 \lambda_0 \tilde{\lambda}_1 y_{2}^2}+\lambda_0 y_{1}+\tilde{\lambda}_1 y_{3}\right).
\end{equation}
Since $\beta\epsilon_c\gg 1$ is assumed, $x>0$ and thus $1/x$ increases as $T$ increases.
Since $1/x_1<1/x_2$, the critical temperature should be determined by $x_2$
and given by
\begin{equation}
\label{Expression_Tc}
T_c=T_0 \exp\left(-\frac{4}{\sqrt{(\lambda_0 y_{1}-\tilde{\lambda}_1 y_3)^2+4 \lambda_0 \tilde{\lambda}_1 y_{2}^2}+\lambda_0 y_{1}+\tilde{\lambda}_1 y_{3}}\right) ,
\end{equation}
where $T_0=2e^{\bar{\gamma}}\epsilon_c/(\pi k_B)$.
The corresponding eigen-vector gives rise to the ratio of order parameters $\frac{\tilde{\Delta}_0}{\tilde{\Delta}_1}$ in different channels, which reads
\begin{equation}
\label{Delta_ratio}
\frac{\tilde{\Delta}_0}{\tilde{\Delta}_1}=\frac{\sqrt{(\lambda_0 y_{1}-\tilde{\lambda}_1 y_3)^2+4 \lambda_0 \tilde{\lambda}_1 y_{2}^2}+\lambda_0 y_{1}-\tilde{\lambda}_1 y_{3}}{2 \tilde{\lambda}_1 y_{2}}.
\end{equation}

We notice that the singlet-quintet mixing can enhance the critical temperature $T_c$. To see that, we can neglect the off-diagonal term in the gap equation (\ref{linearized_gap_eqn_approx}) or equivalently choose $y_2=0$. In this case, the critical temperatures in the singlet and quintet channels can be determined by $1/x_{cs}=\lambda_0 y_1 /2$ and $1/x_{cq}=\tilde{\lambda}_1 y_3 /2$, respectively, where $x_{cs}=\ln(T_0/T_{cs})$ and $x_{cq}=\ln(T_0/T_{cq})$.
Since
\begin{equation}
\frac{1}{x}> \text{max}(\frac{1}{x_{cs}},\frac{1}{x_{cq}})
\end{equation}
with $x=\ln(T_0/T_{c})$, we conclude that the $T_c$ in Eq. (\ref{Expression_Tc}) is always larger than $T_{cs}$ and $T_{cq}$.

\subsection{Kink Structure of $\tilde{\Delta}_1/\tilde{\Delta}_0$ in Regime $III$}
\label{Kink_Regime_III}
This section is devoted to the understanding of three kinks in Fig.1b of the main text, whose positions are shown in Fig.\ref{3half_pairing_c1neqc2_c1c2r}a by gray dashed lines.

Let us first discuss the band structure and the momentum cut-off.
In regime III, the $\xi_-$ bands always bend down, while the $\xi_+$ band bends up along $\Gamma-X$
and down along $\Gamma-L$ or vice verse, as depicted in Fig.1a(iii) in the main text. Therefore, a saddle point exists at $\Gamma$ for the $\xi_+$ bands, and leads to hyperbolic Fermi surface with divergent density of states. Such hyperbolic Fermi surface is due to the limitation of the Luttinger model, which is only valid in a small momentum region around $\Gamma$. More importantly, it will cause the divergence of the functions $y_{1,2,3}$. To avoid this problem, we introduce a momentum cut-off $\Lambda$, which can be implemented by inserting a Heaviside step function
$\theta(\Lambda-\sqrt{2m_-\mu})=\theta(1-2m Q_c-2m\mu/\Lambda^2)$ and $\theta(\Lambda-\sqrt{2m_+\mu})=\theta(1+2m Q_c-2m\mu/\Lambda^2)$
into the integral for the $\xi_-$ and $\xi_+$ bands, respectively. As a result, the functions $y_{1,2,3}$ are re-defined as
$$
y_1\equiv \int \frac{d\Omega}{4\pi}\left[\frac{\theta(1-2m Q_c-2m\mu/\Lambda^2)}{(1-2m Q_c)^{3/2}}+\frac{\theta(1+2m Q_c-2m\mu/\Lambda^2)}{(1+2mQ_c)^{3/2}}\right],
$$
$$
y_2\equiv \int \frac{d\Omega}{4\pi}\left[-\frac{\theta(1-2m Q_c-2m\mu/\Lambda^2)}{(1-2m Q_c)^{5/2}}+\frac{\theta(1+2m Q_c-2m\mu/\Lambda^2)}{(1+2mQ_c)^{5/2}}\right]\frac{|c_1|Q_1^2 + |c_2|Q_2^2}{ Q_c}
$$
and
$$
y_3\equiv \int \frac{d\Omega}{4\pi}\left[\frac{\theta(1-2m Q_c-2m\mu/\Lambda^2)}{(1-2m Q_c)^{7/2}}+\frac{\theta(1+2m Q_c-2m\mu/\Lambda^2)}{(1+2mQ_c)^{7/2}}\right]\left(\frac{Q_1^2  |c_1|+Q_2^2  |c_2|}{ Q_c}\right)^2.
$$

The Fermi surface shape of the $\xi_+$ band plays an important role in determining the values of $y_{1,2,3}$, and consequently the kink structures. 
We choose the momentum cut-off as $\Lambda=3\sqrt{2m\mu}$ for the red line  and $\Lambda=\infty$ for the blue line in Fig. \ref{3half_pairing_c1neqc2_c1c2r} (a). For a small SOC parameter $c_1$, the momentum cut-off is not important and thus the red line coincides with the blue line in Fig. \ref{3half_pairing_c1neqc2_c1c2r} (a). With increasing the SOC to $|2m c_1|=4/9$, the Fermi momentum of the $\xi_+$ band starts becoming larger than the cut-off $\Lambda$ along certain angles, and thus the integrals $y_{1,2,3}$ are limited by $\Lambda$ (see Fig.\ref{3half_pairing_c1neqc2_c1c2r}b), leading to the appearance of the first kink in $\tilde{\Delta}_1/\tilde{\Delta_0}$.
With further increasing $|2m c_1|$, we find $y_{2,3}$ show a peak behavior in Fig.\ref{3half_pairing_c1neqc2_c1c2r}b due to the shrink of the Fermi surface range for the $\xi_+$ bands, giving rise to the second kink. When the SOC reaches $|2m c_1|=8/9$, the $\xi_+$ Fermi surface moves away from the momentum range within the cut-off $\Lambda$. This yields a significant decreasing of $y_{1,2,3}$, as well as a dramatic drop of $\tilde{\Delta}_1/\tilde{\Delta_0}$. The quintet mixing is negligible in the regime III when $|2m c_1|>8/9$.

\begin{figure}[h]
\includegraphics[width=0.5\columnwidth]{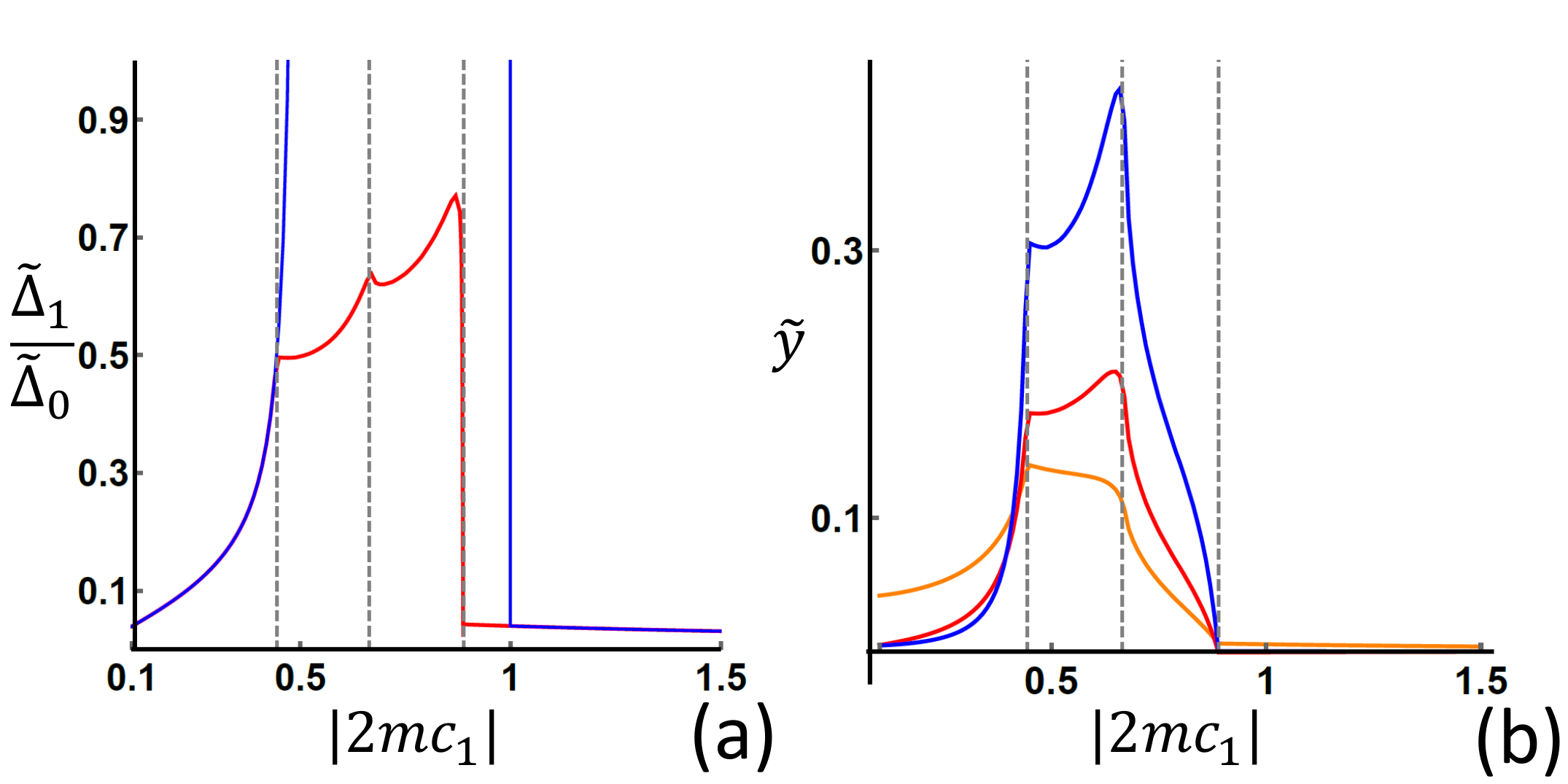}
\caption{
\label{3half_pairing_c1neqc2_c1c2r}
(a) shows the pairing ratio $\tilde{\Delta}_1/\tilde{\Delta}_0$ as a function of the symmetric SOC strength $|2m c_1|$.
The parameter choice is $c_2=2 c_1$ and $\tilde{\lambda}_1/\lambda_0=0.1$.
The momentum cut-off is not considered for the blue lines while $\Lambda=3\sqrt{2m\mu}$ is for the red line.
Three gray dashed lines mark the position of three kinks of the red lines.
(b) shows $\tilde{y}_1=\lambda_0 y_1$ (orange), $\tilde{y}_2=\sqrt{\lambda_0\tilde{\lambda}_1} y_2$ (red) and $\tilde{y}_3=\tilde{\lambda}_1 y_3$(blue) as a function of the symmetric SOC strength $|2m c_1|$ with $c_2=2 c_1$, $\tilde{\lambda}_1/\lambda_0=0.1$, $\lambda_0=0.02$ and $\Lambda=3\sqrt{2m\mu}$.
The three gray dashed lines are at the same positions of those in (a), standing for the positions of three kinks.
The negative chemical potential is used as a unit and does not need a specific value.}
\end{figure}

\section{Bogoliubov-de Gennes Hamiltonian and Topological Nodal-line superconductivity}
We will study the energy dispersion of the Bogoliubov-de Gennes (BdG) Hamiltonian with the singlet-quintet mixing and extract the phase diagram for the topological nodal-line superconducting phase.

\subsection{BdG Hamiltonian}
Here we first give a review of the BdG Hamiltonian for superconductivity and its symmetry property.
The BdG Hamiltonian is written as
\begin{equation}
H=\frac{1}{2}\sum_{\mathbf{k}}\Psi^{\dagger}_{\mathbf{k}}h_{BdG}(\mathbf{k})\Psi_{\mathbf{k}}+\text{const.}
\end{equation}
with
\begin{equation}
\label{h_BdG}
h_{BdG}(\mathbf{k})=
\left(
\begin{matrix}
h(\mathbf{k})& \Delta(\mathbf{k})\\
\Delta^{\dagger}(\mathbf{k})& -h^T(-\mathbf{k})\\
\end{matrix}
\right)
\end{equation}
and $\Psi^{\dagger}_{\mathbf{k}}=(c_{\mathbf{k}}^{\dagger},c_{-\mathbf{k}}^{T})$.

The BdG Hamiltonian has particle-hole symmetry, time reversal symmetry and consequently chiral symmetry.
The particle-hole symmetry is defined as
\begin{equation}
-\mathcal{C}h_{BdG}^*(-\mathbf{k})\mathcal{C}^{\dagger}=h_{BdG}(\mathbf{k})
\end{equation}
with
\begin{equation}
\mathcal{C}=
\left(
\begin{matrix}
0 & \mathds{1}_4\\
\mathds{1}_4 & 0\\
\end{matrix}
\right).
\end{equation}
The time-reversal symmetry is defined as
\begin{equation}
\mathcal{T}h_{BdG}^*(-\mathbf{k})\mathcal{T}^{\dagger}=h_{BdG}(\mathbf{k})
\end{equation}
with
\begin{equation}
\mathcal{T}=
\left(
\begin{matrix}
\gamma & 0\\
0 & \gamma^*\\
\end{matrix}
\right)
=
\left(
\begin{matrix}
\gamma & 0\\
0 & -\gamma^{\dagger}\\
\end{matrix}
\right)
\end{equation}
and $\gamma$ is defined in Sec.A.
From the above definition, we find the requirement
$\Delta(\mathbf{k})\gamma^{\dagger}=\gamma \Delta^{\dagger}(\mathbf{k})$
for the gap function $\Delta(\mathbf{k})$.
We know the BdG Hamiltonian has time-reversal symmetry is because
$\Delta_{0}/\Delta_{1}$ in Eq.\ref{Delta_ratio} is a real number.
According to the convention we choose for time-reversal operator $\gamma=-i\Gamma^{13}$,
$\Delta_{0}$ and $\Delta_{1}$ should be set to be real.

The chiral symmetry is given by
\begin{equation}
\mathcal{\chi}h_{BdG}(\mathbf{k})\mathcal{\chi}^{\dagger}=-h_{BdG}(\mathbf{k}),
\end{equation}
with the chiral operator $\mathcal{\chi}=i\mathcal{T}\mathcal{C}^*$ naturally following
the definition of $\mathcal{T}$ and $\mathcal{C}$.

We can introduce the unitary transformation matrix $U_{\mathcal{\chi}}$
to diagonalize the chiral operator \cite{PhysRevB.81.134508}
\begin{equation}
U_{\mathcal{\chi}}\mathcal{\chi}U_{\mathcal{\chi}}^{\dagger}
=
\left(
\begin{matrix}
-\mathds{1}_4 & 0\\
0 &  \mathds{1}_4\\
\end{matrix}
\right)
\ \text{for}\
U_{\mathcal{\chi}}
=
\frac{1}{\sqrt{2}}
\left(
\begin{matrix}
\mathds{1}_4 & -i\gamma\\
\mathds{1}_4 & i\gamma\\
\end{matrix}
\right).
\end{equation}
Correspondingly, the BdG Hamiltonian can be transformed into an off-diagonal form
\begin{equation}
\label{offdiagnol_HBdG}
U_{\mathcal{\chi}}h_{BdG}(\mathbf{k})U_{\mathcal{\chi}}^{\dagger}
=
\left(
\begin{matrix}
 & h(\mathbf{k})-i \Delta(\mathbf{k})\gamma^{\dagger}\\
h(\mathbf{k})+i \Delta(\mathbf{k})\gamma^{\dagger}& \\
\end{matrix}
\right)
\end{equation}
by the unitary transformation matrix $U_{\mathcal{\chi}}$.

\subsection{Conditions of Nodal lines}
Now we want to extract the conditions for the existence of nodal points or lines in the above BdG Hamiltonian.
Due to chiral symmetry, the energy of nodal points must be zero, thus requiring the condition $\text{det}[h_{BdG}(\mathbf{k})]=0$.
From the Eq.\ref{offdiagnol_HBdG}, in which $h(\mathbf{k})$ is of even dimension, we obtain
\begin{equation}
\text{det}[h_{BdG}(\mathbf{k})]=\text{det}[h(\mathbf{k})-i \Delta(\mathbf{k})\gamma^{\dagger}]\text{det}[h(\mathbf{k})+i \Delta(\mathbf{k})\gamma^{\dagger}]=|\text{det}[h(\mathbf{k})-i \Delta(\mathbf{k})\gamma^{\dagger}]|^2,
\end{equation}
and thus
\begin{equation}
\text{det}[h_{BdG}(\mathbf{k})]=0\Leftrightarrow Re\{\text{det}[h(\mathbf{k})-i \Delta(\mathbf{k})\gamma^{\dagger}]\}=0\ \&\  Im\{\text{det}[h(\mathbf{k})-i \Delta(\mathbf{k})\gamma^{\dagger}]\}=0.
\end{equation}

According to the Luttinger model expression and gap function expressions (\ref{paring_form}), we have
\begin{equation}
\label{eq_det}
\text{det}[h(\mathbf{k})-i \Delta(\mathbf{k})\gamma^{\dagger}]=\frac{1}{16} \left(\Delta_{0}^2-a^4 \Delta_{1}^2 k^4+4 k^4 Q_{c}^2-4 \xi ^2+4 i \left(\Delta_{0} \xi -a^2 \Delta_{1} k^4 \left(c_{1}  Q_{1}^2+c_{2} Q_{2}^2\right)\right)\right)^2,
\end{equation}
which leads to
\begin{equation}
\Delta_{0}^2-a^4 \Delta_{1}^2 k^4+4 k^4 Q_c^2-4 \xi ^2=0\ \&\ \Delta_{0} \xi -a^2 \Delta_{1}k^4 \left(c_{1}  Q_{1}^2+c_{2} Q_{2}^2\right)=0.
\end{equation}

Under the conditions $m<0$,$\mu<0$ and $c_1c_2>0$, the above equations can be simplified as
\begin{equation}
\label{exact_nodal_condition}
\left(\frac{\tilde{\Delta}_{0}}{\mu}\right)^2- \left(\frac{\tilde{\Delta}_{1}}{\mu}\right)^2 \tilde{k}^4+4 \tilde{k}^4 (2m Q_c)^2-4 (\tilde{k}^2-1) ^2=0\ \&\ \frac{\tilde{\Delta}_{0}}{\tilde{\Delta}_{1}} (\tilde{k}^2-1) -\tilde{k}^4 \left(2m|c_{1}|  Q_{1}^2+2m|c_{2}| Q_{2}^2\right)=0
\end{equation}
with $\tilde{k}=k/\sqrt{ 2 m \mu }$.
One can numerically solve the above equations for
$\tilde{k}^2$ and $Q_1^2$ with $Q_2^2=1-Q_1^2$ and $Q_c^2=c_1^2 Q_1^2+c_2^2 Q_2^2$
to extract the existence and location of nodal points or lines.

Below we will further demonstrate the 4-fold degeneracy at each nodal point for the BdG Hamiltonian of the Luttinger model. This is due to inversion symmetry, which is given by
\begin{equation}
\mathcal{P}h_{BdG}(-\mathbf{k})\mathcal{P}^{\dagger}=h_{BdG}(\mathbf{k})
\end{equation}
with $\mathcal{P}=-\mathds{1}_{8\times 8}$, in addition to Time reversal, particle-hole symmetry
and chiral symmetry. By combining inversion with time-reversal or particle-hole, we can define two new symmetry operators: $\widetilde{\mathcal{T}}=\mathcal{P}\mathcal{T}$ and $\widetilde{\mathcal{C}}=\mathcal{P}\mathcal{C}$, which satisfy the symmetry relations
\begin{equation}
\widetilde{\mathcal{T}}h_{BdG}^*(\mathbf{k})\widetilde{\mathcal{T}}^{\dagger}=\mathcal{P}\mathcal{T}h_{BdG}^*(\mathbf{k})\mathcal{T}^{\dagger}\mathcal{P}^{\dagger}=h_{BdG}(\mathbf{k})
\end{equation}
and
\begin{equation}
-\widetilde{\mathcal{C}}h_{BdG}^*(\mathbf{k})\widetilde{\mathcal{C}}^{\dagger}=-\mathcal{P}\mathcal{C}h_{BdG}^*(\mathbf{k})\mathcal{C}^{\dagger}\mathcal{P}^{\dagger}=h_{BdG}(\mathbf{k}).
\end{equation}
Since the momentum ${\bf k}$ is invariant under $\widetilde{T}$ and $\widetilde{C}$, we conclude that any nodal point at zero energy should be 4-fold degenerate.

\subsection{Projection of gap function onto the Fermi surface}
Although the nodal points or lines can be determined by Eq.\ref{exact_nodal_condition} numerically, it is desirable to have more analytic understanding of the origin of these nodal points or lines. In this section, we will project the gap function onto the Fermi surface, from which one can identify the physical origin of the nodal points and lines.
The BdG Hamiltonian (\ref{h_BdG}) can be re-written in a compact form as
\begin{equation}
h_{BdG}(\mathbf{k})=\frac{\tau_0+\tau_3}{2}h(\mathbf{k})+\frac{\tau_0-\tau_3}{2}(-h^T(-\mathbf{k}))+\frac{\tau_+}{2}\Delta(\mathbf{k})+\frac{\tau_-}{2}\Delta^{\dagger}(\mathbf{k}),
\end{equation}
where $\tau_0$ and $\tau_{1,2,3}$ are identity matrix and Pauli matrices for particle hole index and $\tau_{\pm}=\tau_1\pm i\tau_2$. The Luttinger Hamiltonian $h({\bf k})$ can be diagonalized by the unitary transformation
\cite{PhysRevB.69.235206}
$$
U(\mathbf{k})\equiv \frac{1}{\sqrt{2(1+\frac{c_2 g_{\mathbf{k},5}}{|c_1|\tilde{g}_{\mathbf{k}}})}}\left((1+\frac{c_2 g_{\mathbf{k},5}}{|c_1|\tilde{g}_{\mathbf{k}}})\Gamma^0 +i\sum_{a=1}^4\frac{c_1 \tilde{g}_{\mathbf{k},a}}{|c_1|\tilde{g}_{\mathbf{k}}}\Gamma^{a5} \right)D,
$$
with
$$
D=\left(
\begin{array}{cccc}
 1 & 0 & 0 & 0 \\
 0 & 0 & 1 & 0 \\
 0 & 0 & 0 & 1 \\
 0 & 1 & 0 & 0 \\
\end{array}
\right).
$$
This leads to
$$
U^{\dagger}(\mathbf{k})h(\mathbf{k})U(\mathbf{k})=D^{\dagger} (\xi_{\mathbf{k}} \Gamma^0+Q_c k^2\Gamma^5) D=\xi_{\mathbf{k}} \Gamma^0-Q_c k^2\Gamma^{34}=
\left(
\begin{array}{cccc}
 \xi_+ & 0 & 0 & 0 \\
 0 & \xi_+ & 0 & 0 \\
 0 & 0 & \xi_- & 0 \\
 0 & 0 & 0 & \xi_- \\
\end{array}
\right).
$$

We define the unitary transformation
$$
U_{BdG}(\mathbf{k})
\equiv
\frac{\tau_0+\tau_3}{2}U(\mathbf{k})+\frac{\tau_0-\tau_3}{2}\gamma^{\dagger} U(\mathbf{k}),
$$
which gives rise to
$$
U_{BdG}^{\dagger}(\mathbf{k})h_{BdG}(\mathbf{k})U_{BdG}(\mathbf{k})
=
$$
$$
\frac{\tau_0+\tau_3}{2}[\xi_{\mathbf{k}} \Gamma^0-Q_c k^2\Gamma^{34}]
+ \frac{\tau_0-\tau_3}{2}[-\xi_{\mathbf{k}} \Gamma^0+Q_c k^2\Gamma^{34}]
 +  \left[\frac{\tau_+}{2}U^{\dagger}(\mathbf{k})\Delta(\mathbf{k})\gamma^{\dagger} U(\mathbf{k})+h.c.\right]
$$

With the pairing expression (\ref{paring_form}),
we can obtain
$$
U^{\dagger}(\mathbf{k})\Delta(\mathbf{k})\gamma^{\dagger} U(\mathbf{k})
=
$$
\begin{equation}
\label{BdG_pairing_term}
\left(
\begin{array}{cccc}
\frac{\Delta_0}{2}+\frac{\Delta_1}{2}k^2 a^2\frac{c_1Q_1^2+c_2 Q_2^2}{Q_c}
&
0
&
\frac{k^2a^2 \Delta_1(c_1-c_2)}{Q_c^2+c_2 \hat{g}_{\mathbf{k},5}Q_c}i f_{1}(\hat{\mathbf{k}})
&
\frac{k^2a^2 \Delta_1(c_1-c_2)}{Q_c^2+c_2 \hat{g}_{\mathbf{k},5}Q_c}f_{2}(\hat{\mathbf{k}})
\\
 0
 &
 \frac{\Delta_0}{2}+\frac{\Delta_1}{2}k^2 a^2\frac{c_1Q_1^2+c_2 Q_2^2}{Q_c}
 &
 \frac{k^2a^2 \Delta_1(c_1-c_2)}{Q_c^2+c_2 \hat{g}_{\mathbf{k},5}Q_c}f_{2}^*(\hat{\mathbf{k}})
 &
 \frac{k^2a^2 \Delta_1(c_1-c_2)}{Q_c^2+c_2 \hat{g}_{\mathbf{k},5}Q_c}i f_{1}^*(\hat{\mathbf{k}})
 \\
\frac{k^2a^2 \Delta_1(c_1-c_2)}{Q_c^2+c_2 \hat{g}_{\mathbf{k},5}Q_c}(-i) f_{1}^*(\hat{\mathbf{k}})
 &
 \frac{k^2a^2 \Delta_1(c_1-c_2)}{Q_c^2+c_2 \hat{g}_{\mathbf{k},5}Q_c}f_{2}(\hat{\mathbf{k}})
 &
 \frac{\Delta_0}{2}-\frac{\Delta_1}{2}k^2 a^2\frac{c_1Q_1^2+c_2 Q_2^2}{Q_c}
 &
 0
 \\
 \frac{k^2a^2 \Delta_1(c_1-c_2)}{Q_c^2+c_2 \hat{g}_{\mathbf{k},5}Q_c}f_{2}^*(\hat{\mathbf{k}})
 &
\frac{k^2a^2 \Delta_1(c_1-c_2)}{Q_c^2+c_2 \hat{g}_{\mathbf{k},5}Q_c}(-i) f_{1}(\hat{\mathbf{k}})
 &
 0
 &
 \frac{\Delta_0}{2}-\frac{\Delta_1}{2}k^2 a^2\frac{c_1Q_1^2+c_2 Q_2^2}{Q_c}
 \\
\end{array}
\right),
\end{equation}
where $2 f_1(\hat{\mathbf{k}})=(\hat{g}_{\mathbf{k},1}+i \hat{g}_{\mathbf{k},2})(c_2 Q_2^2+\hat{g}_{\mathbf{k},5}Q_c)$ and $2 f_2(\hat{\mathbf{k}})=c_1Q_1^2\hat{g}_{\mathbf{k},4}+i \hat{g}_{\mathbf{k},3}(c_2 Q_2^2+\hat{g}_{\mathbf{k},5}Q_c)$.

After the unitary transformation,
the block part of the $\xi_+$ bands in the BdG Hamiltonian is given by
$$
\left(
\begin{array}{cccc}
\frac{1}{2m_+}k^2-\mu & \frac{\Delta_0}{2}+\frac{\Delta_1}{2}k^2 a^2\frac{c_1Q_1^2+c_2 Q_2^2}{Q_c}& 0 & 0\\
 \frac{\Delta_0^*}{2}+\frac{\Delta_1^*}{2}k^2 a^2\frac{c_1Q_1^2+c_2 Q_2^2}{Q_c} &-\frac{1}{2m_+}k^2+\mu  & 0 & 0\\
0 &0 &\frac{1}{2m_+}k^2-\mu & \frac{\Delta_0}{2}+\frac{\Delta_1}{2}k^2 a^2\frac{c_1Q_1^2+c_2 Q_2^2}{Q_c}\\
0 &0 & \frac{\Delta_0^*}{2}+\frac{\Delta_1^*}{2}k^2 a^2\frac{c_1Q_1^2+c_2 Q_2^2}{Q_c} &-\frac{1}{2m_+}k^2+\mu \\
\end{array}
\right)
$$
while the block for the $\xi_-$ bands is
$$
\left(
\begin{array}{cccc}
\frac{1}{2m_-}k^2-\mu & \frac{\Delta_0}{2}-\frac{\Delta_1}{2}k^2 a^2\frac{c_1Q_1^2+c_2 Q_2^2}{Q_c}& 0 & 0\\
 \frac{\Delta_0^*}{2}-\frac{\Delta_1^*}{2}k^2 a^2\frac{c_1Q_1^2+c_2 Q_2^2}{Q_c} &-\frac{1}{2m_-}k^2+\mu  & 0 & 0\\
0 &0 &\frac{1}{2m_-}k^2-\mu & \frac{\Delta_0}{2}-\frac{\Delta_1}{2}k^2 a^2\frac{c_1Q_1^2+c_2 Q_2^2}{Q_c}\\
0 &0 & \frac{\Delta_0^*}{2}-\frac{\Delta_1^*}{2}k^2 a^2\frac{c_1Q_1^2+c_2 Q_2^2}{Q_c} &-\frac{1}{2m_-}k^2+\mu \\
\end{array}
\right).
$$
The coupling between different $\xi_{\pm}$ blocks is given by the off-diagonal terms of Eq. (\ref{BdG_pairing_term}), which is zero in the isotropic limit($c_1=c_2$) and can be neglected for small anisotropy.
Even if the anisotropy is not small, it can still be neglected since the physics related with pairing is only relevant near Fermi surfaces.

From the expressions of $\xi_{\pm}$ blocks, we notice that the d-wave quintet pairing is transformed into s-wave singlet pairing with $k^2$ dependence after the projection. Such form of pairing is normally known as extended s-wave pairing in literature\cite{PhysRevB.64.024503,PhysRevB.65.054503}. As a result, it is easy to see that the nodal condition is determined by the vanishing of this extended s-wave gap function at the Fermi surface of each band. This Fermi surface project scheme provides a more clear physical picture of how the singlet-quintet mixing mechanism can induce nodal points or lines in the gap function.

For the $\xi_+$ band, the nodal condition is determined by
$$
\frac{1}{2m_+}k^2-\mu=0\ \text{and} \ \ \frac{\Delta_0}{2}+\frac{\Delta_1}{2}k^2 a^2\frac{c_1Q_1^2+c_2 Q_2^2}{Q_c}=0,
$$
which gives
\begin{equation}
\label{nodal_condition_E+_without_lieqn}
1+2m Q_c > 0 \ \&\ k=\sqrt{2 m_+ \mu} \ \&\ \frac{\tilde{\Delta}_0}{\tilde{\Delta}_1}=-\frac{1}{1+2m Q_c}\frac{|c_1|Q_1^2+|c_2| Q_2^2}{Q_c}.
\end{equation}
For the $\xi_-$ band, the nodal condition is
$$
\frac{1}{2m_-}k^2-\mu=0\ \text{and} \ \ \frac{\Delta_0}{2}-\frac{\Delta_1}{2}k^2 a^2\frac{c_1Q_1^2+c_2 Q_2^2}{Q_c}=0,
$$
which gives
\begin{equation}
\label{nodal_condition_E-_without_lieqn}
1-2m Q_c > 0 \ \&\ k=\sqrt{2 m_- \mu} \ \& \ \frac{\tilde{\Delta}_0}{\tilde{\Delta}_1}= \frac{1}{1-2 m Q_c}\frac{|c_1|Q_1^2+|c_2| Q_2^2}{Q_c}.
\end{equation}
Here we have used $m<0$, $\mu <0$ and $c_1 c_2 >0$ and the relation between $\tilde{\Delta}_0$ ($\tilde{\Delta}_1$) and $\Delta_0$ ($\Delta_1$) is defined in the main text.



The nodal condition for the $\xi_+$ ($\xi_-$) band requires
$\tilde{\Delta}_0/\tilde{\Delta}_1<0$ ($\tilde{\Delta}_0/\tilde{\Delta}_1>0$).
According to Eq. (\ref{Delta_ratio}), which is solved from the linearized gap equation,
the sign of $\tilde{\Delta}_0/\tilde{\Delta}_1$ is determined by $y_2$.
Therefore, we need to discuss two cases with different signs of $y_2$, separately.

If $y_2>0$, $\tilde{\Delta}_0/\tilde{\Delta}_1>0$ and thus nodal points cannot exist on the $\xi_+$ Fermi surface. In this case, the nodal points for the $\xi_-$ band require
\begin{equation}
\label{nodal_condition_E-}
1-2m Q_c > 0 \ \&\ k=\sqrt{2 m_- \mu} \ \& \ \frac{\tilde{\lambda}_1}{\lambda_0}= \frac{f_- y_1+y_2}{f_-^2 y_2+f_- y_3}\ \&\ \ y_2>0,
\end{equation}
where $f_-=\frac{1}{1-2 m Q_c}\frac{|c_1|Q_1^2+|c_2| Q_2^2}{Q_c}$.

If $y_2<0$, $\tilde{\Delta}_0/\tilde{\Delta}_1<0$ and nodal points cannot exist on the $\xi_-$ Fermi surface.
The nodal points for $\xi_+$ bands are fixed by
\begin{equation}
\label{nodal_condition_E+}
1+2m Q_c > 0 \ \&\ k=\sqrt{2 m_+ \mu} \ \&\ \frac{\tilde{\lambda}_1}{\lambda_0}= \frac{f_+ y_1-y_2}{-f_+^2 y_2+f_+ y_3}\ \&\ \ y_2<0,
\end{equation}
where $f_+=\frac{1}{1+2m Q_c}\frac{|c_1|Q_1^2+|c_2| Q_2^2}{Q_c}$.
The nodal lines extracted from the above equations are found to fit well with those obtained from
the direct numerical calculations of Eq. (\ref{exact_nodal_condition}) for $\Delta_1$ being not too large.

\subsection{Topological invariant for Nodal Lines}
In this section, we will extract topological nature of nodal lines in the phase diagram by defining appropriate topological invariants. Due to the existence of chiral symmetry, the BdG Hamiltonian at an arbitrary momentum ${\bf k}$ belongs to the AIII class. Thus we consider the one dimentional topological invariant $N_{w}$ in the AIII class, defined as \cite{PhysRevB.84.060504}
\begin{equation}
\label{1dAIII_TI}
N_{w}=\frac{1}{2\pi i}\int_{\mathcal{L}} d\mathbf{k}\cdot \text{Tr}[Q^{\dagger}(\mathbf{k})\mathbf{\nabla}_{\mathbf{k}}Q(\mathbf{k})],
\end{equation}
where $\mathcal{L}$ is chosen to be a closed path that does not pass any gapless point in the momentum space.
Here the matrix $Q(\mathbf{k})=U(\mathbf{k})V^{\dagger}(\mathbf{k})$, in which
$U$ and $V$ are two unitary matrices from the singular value decomposition of the upper off-diagonal block of transformed Hamiltonian in Eq.\ref{offdiagnol_HBdG},
\begin{equation}
h(\mathbf{k})-i \Delta(\mathbf{k})\gamma^{\dagger}=U(\mathbf{k})\Sigma(\mathbf{k})V^{\dagger}(\mathbf{k})
\end{equation}
and $\Sigma(\mathbf{k})$ is a diagonal matrix with entries being real and non-negative.
One can easily show that $Q$ is unitary $Q^{\dagger}=Q^{-1}$ and
\begin{equation}
\text{Tr}[Q^{\dagger}(\mathbf{k})\mathbf{\nabla}_{\mathbf{k}}Q(\mathbf{k})]
=
\text{Tr}[Q^{-1}(\mathbf{k})\mathbf{\nabla}_{\mathbf{k}}Q(\mathbf{k})]
=
\text{Tr}[\mathbf{\nabla}_{\mathbf{k}}\text{ln} (Q(\mathbf{k}))]
=
\mathbf{\nabla}_{\mathbf{k}}\text{Tr}[\text{ln} (Q(\mathbf{k}))]
=
\mathbf{\nabla}_{\mathbf{k}}\text{ln}[\text{Det}(Q(\mathbf{k}))]
=
\mathbf{\nabla}_{\mathbf{k}}i\text{Arg}[\text{Det}(Q(\mathbf{k}))],
\end{equation}
where $|\text{Det}[Q(\mathbf{k})]|=1$ is used in the last equality and $\text{Arg}[x]$ is defined as $x=|x|e^{i\text{Arg}[x]}$.
With these derivations, we obtain
\begin{equation}
\label{N_Arg_Q}
N_{w}=\frac{1}{2\pi }\int_{\mathcal{L}} d\mathbf{k}\cdot \mathbf{\nabla}_{\mathbf{k}}\text{Arg}[\text{Det}(Q(\mathbf{k}))],
\end{equation}
which means $N_{w}\in\mathds{Z}$.
On the other hand, since $U$, $\Sigma$ and $V$ are square matrix, we have
\begin{equation}
\text{Det}[h(\mathbf{k})-i \Delta(\mathbf{k})\gamma^{\dagger}]
=
\text{Det}[U\Sigma V^{\dagger}]
=
\text{Det}[U]\text{Det}[\Sigma]\text{Det}[V^{\dagger}]
=
\text{Det}[Q]\text{Det}[\Sigma].
\end{equation}
Since the eigen spectrum of the BdG Hamiltonian along $\mathcal{L}$ is gaped, we have $\text{Det}[\Sigma]> 0$ and
\begin{equation}
\text{Det}[Q]=\text{Det}[h(\mathbf{k})-i \Delta(\mathbf{k})\gamma^{\dagger}]/\text{Det}[\Sigma]
\Rightarrow
\text{Arg}[\text{Det}[Q]]=\text{Arg}[\text{Det}(h(\mathbf{k})-i \Delta(\mathbf{k})\gamma^{\dagger})].
\end{equation}
This derivation eventually leads to
\begin{equation}
\label{N_Arg_h}
N_{w}=\frac{1}{2\pi }\int_{\mathcal{L}} d\mathbf{k}\cdot \mathbf{\nabla}_{\mathbf{k}}\text{Arg}[\text{Det}(h(\mathbf{k})-i \Delta(\mathbf{k})\gamma^{\dagger})],
\end{equation}
from which one can see that the physical meaning of topological invariant $N_w$ is the winding number of the quantity $\text{Det}(h(\mathbf{k})-i \Delta(\mathbf{k})\gamma^{\dagger})$ along the closed path $\mathcal{L}$.
We apply the above formula to Eq.\ref{eq_det} for the BdG Hamiltonian of the Luttinger model with singlet-quintet mixing and find that all nodal lines carry a non-trivial topological invariant
\begin{equation}
N_{w}=\pm 2.
\end{equation}
The even number of $N_w$ coincides with the 4-fold degeneracy at each point along the nodal line.

\subsection{Topological Nodal-line Superconductivity of the Luttinger model}
In this section we will discuss the possibility of topological nodal-line superconductivity for the Luttinger model in different parameter regimes (Regime I, II and III).

\subsubsection{Regime I: Normal Band Structure}
\label{Attractive_Regime_I_Nodal}
We first consider the isotropic case ($c_1=c_2$) for the regime I, in which the condition $0>2m Q_c>-1$ is satisfied. In this case, $Q_c=|c_1|$ and $|c_1| Q_1^2+ |c_2| Q_2^2=|c_1|$. Thus, the condition $2m Q_c>-1$ can be simplified as $2m |c_1|>-1$. With $1+2m|c_1|>0$ and $1-2m |c_1|>0$ ($m<0$), the functions $y_{1,2,3}$ are simplified as
$$
y_1=\frac{1}{(1-2m |c_1|)^{3/2}}+\frac{1}{(1+2m|c_1|)^{3/2}},
$$
$$
y_2= -\frac{1}{(1-2m |c_1|)^{5/2}}+\frac{1}{(1+2m |c_1|)^{5/2}},
$$
and
$$
y_3= \frac{1}{(1-2m |c_1|)^{7/2}}+\frac{1}{(1+2m|c_1|)^{7/2}}.
$$

Nodal points or lines can not exist on the Fermi surface of $\xi_+$ bands due to $y_2>0$. Given $c_1=c_2$ and $f_-=1/(1-2m|c_1|)$, the nodal condition Eq. (\ref{nodal_condition_E-}) is simplified as
\begin{equation}
k=\sqrt{2 m_- \mu} \ \& \ \frac{\tilde{\lambda}_1}{\lambda_0}= 1-(2m|c_1|)^2
\end{equation}
for the $\xi_-$ bands.
Due to the isotropy of the model, the whole Fermi surface of the $\xi_-$ bands will become nodal in this case.

The above discussion of the isotropic case can be easily generalized to the anisotropic case ($c_1\neq c_2$). However, since the integral in $y_{1,2,3}$ cannot be evaluated analytically in this case, we can only solve the nodal condition Eq. (\ref{nodal_condition_E-}) numerically. The qualitative conclusion from the isotropic case still exists in the anisotropic case. We expect the nodal points can only exist on the Fermi surface of the $\xi_-$ bands, but not on that of the $\xi_+$ bands due to positive $y_2$. However, since the Fermi surface is anisotropic in this case, the nodal condition Eq. (\ref{nodal_condition_E-}) is only satisfied at certain angle, leading to the nodal rings, as shown in the Fig. 2 in the main text.

\subsubsection{Regime II: Inverted Band Structure}
\label{Attractive_Regime_II_Nodal}
In this parameter regime, we find $2m Q_c<-1$, and thus $\sqrt{1+2m Q_c}$ is purely imaginary, leading to $y_2<0$ according to Eq. (\ref{eq_y2}).
The nodal conditions Eq. (\ref{nodal_condition_E+}) and (\ref{nodal_condition_E-}) for the $\xi_+$
and $\xi_-$ bands both cannot be satisfied for $2m Q_c<-1$ and $y_2<0$.
Therefore, no nodal lines can exist in this case. It should be mentioned that if interaction in the quintet channel is repulsive, instead of attractive, nodal lines are still possible and we will discuss this situation in the later section.

\subsubsection{Regime III: A special type of inverted band structure with saddle point}
\label{Attractive_Regime_III_Nodal}
\begin{figure}[h]
\includegraphics[width=0.9\columnwidth]{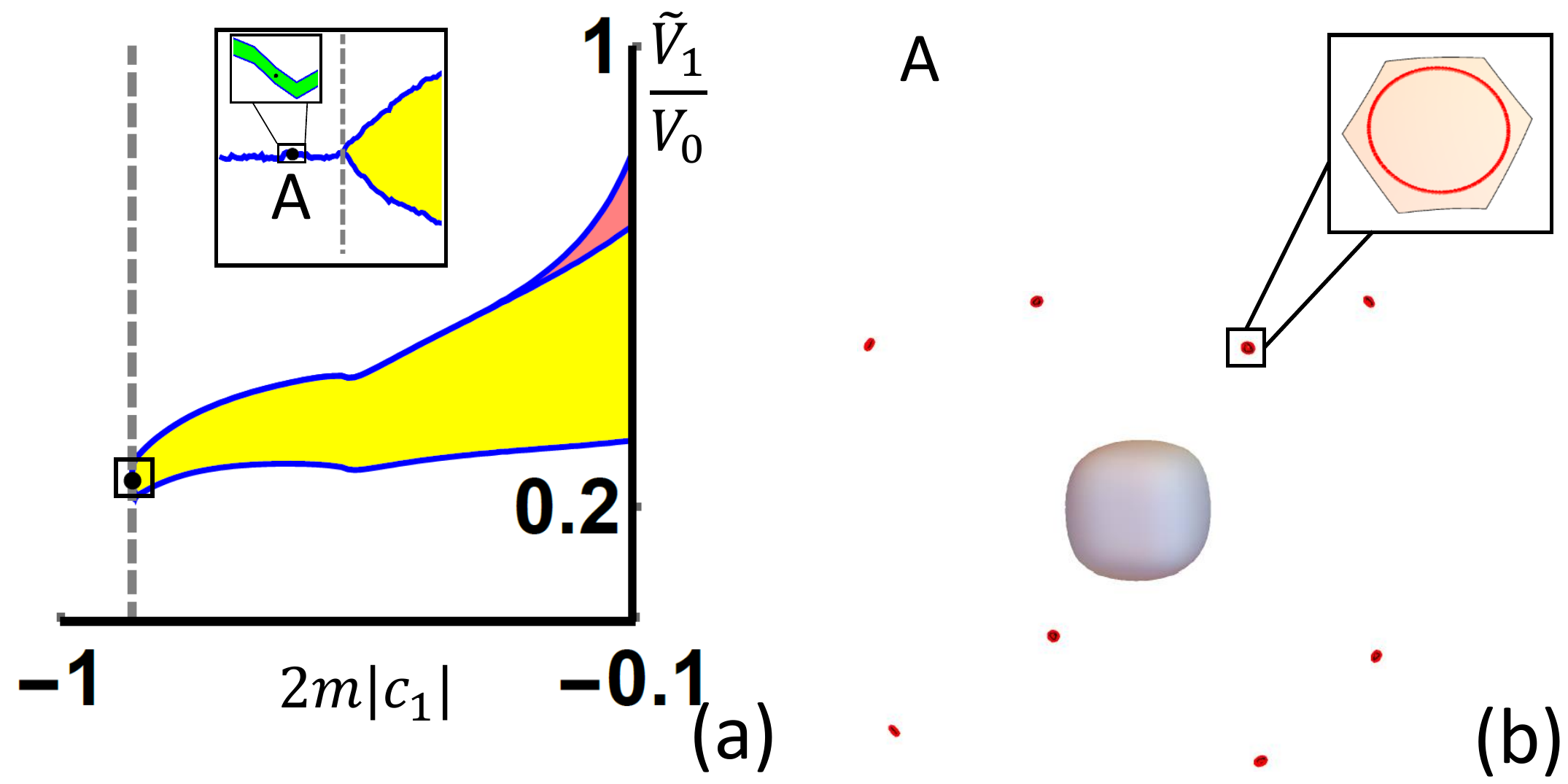}
\caption{
\label{3half_pairing_c1neqc2_Divergent_Fermi_Surface}
(a) shows a phase diagram in $(2m|c_1|,\tilde{V}_1/V_0)$ space for $2m|c_2|=-1.5$ and $\Lambda/\sqrt{2m\mu}=3$.
In the yellow(red) region, nodal lines exist on the $\xi_-$ Fermi surface and the corresponding distribution is similar as Fig.2b(e) of the main text, while
nodal lines can only exist on $\xi_+$ Fermi surface on the left of the dashed line.
The parameter region for existence of nodal lines on $\xi_+$ surface inside the momentum cut-off $\Lambda$ is very narrow (looking like a line even in the inset) and point A is inside the region (inset).
In the inset of the inset, nodal lines exist on $\xi_+$ surface inside the momentum cut-off $\Lambda$ in the green region and the point is the A point.
Parameter choice for point A is $2m|c_1|=-0.8886$, $2m|c_2|=-1.5$, $\Lambda/\sqrt{2m\mu}=3$ and $\tilde{\lambda}_1/\lambda_0=0.2434$.
The dashed line is around  $2m|c_1|=-0.888395$.
(b)
 shows the distribution of nodal lines(red circles) on the $\xi_+$ Fermi surface  inside $\Lambda$ for point A in (a). The outer surfaces are very small since only the part inside momentum cutoff is plotted and the zoom-in version is shown in the inset.
}
\end{figure}

Fig.\ref{3half_pairing_c1neqc2_Divergent_Fermi_Surface}a shows a phase diagram in the parameter space of $2m|c_1|$ and $\tilde{V}_1/V_0)$ for $2m|c_2|=-1.5$ and $\Lambda/\sqrt{2m\mu}=3$ in the regime III.
In the yellow and red regions of the phase diagram, nodal lines exist on the Fermi surface of the $\xi_-$ bands. This nodal phase is quite similar as that in the regime I and has been well discussed in Fig.2b(e) of the main text.

Here we focus on the possibility of nodal rings on the Fermi surface of the $\xi_+$ bands.
Due to the momentum cut-off $\Lambda$, $y_2$ can be negative if only a sufficiently small area of $\xi_+$ Fermi surface is included, and thus nodal lines can also exist on the $\xi_+$ Fermi surface within $\Lambda$, according to the nodal condition Eq.\ref{nodal_condition_E+}.
The corresponding region is on the left of the dashed line Fig.\ref{3half_pairing_c1neqc2_Divergent_Fermi_Surface}a in the phase diagram and very narrow (green region in the inset of inset of Fig.\ref{3half_pairing_c1neqc2_Divergent_Fermi_Surface}a).
Fig.\ref{3half_pairing_c1neqc2_Divergent_Fermi_Surface}b shows that eight nodal rings exist on the Fermi surface of the $\xi_+$ bands at the point A in the green region of Fig.\ref{3half_pairing_c1neqc2_Divergent_Fermi_Surface}a.

\subsection{Nodal Superconductivity with Inversion Breaking Term}
We have neglected the small inversion breaking term (anti-symmetric SOC term $\sim 0.01eV$) in the main text. In this section, we will include this term and show its influence on nodal-line superconductivity.
The leading order of the anti-symmetric SOC has the following form
\begin{equation}
h_{IB}(\mathbf{k})=\frac{2}{\sqrt{3}}C (k_x V_x+ k_y V_y+ k_z V_z),
\end{equation}
where $V_x=\frac{1}{2}\{J_x,J_y^2-J_z^2\}$,
$V_y=\frac{1}{2}\{J_y,J_z^2-J_x^2\}$ and
$V_z=\frac{1}{2}\{J_z,J_x^2-J_y^2\}$.
In terms of spin tensors, three $V$s can be re-written as
\begin{equation}
\left\{
\begin{array}{l}
V_x=\frac{\sqrt{15}}{4}(M^{31}-M^{3,-1})+\frac{3}{4}(M^{33}-M^{3,-3})\\
V_y=i\frac{\sqrt{15}}{4}(M^{31}+M^{3,-1})-i\frac{3}{4}(M^{33}+M^{3,-3})\\
V_z=\sqrt{\frac{3}{2}}(M^{32}+M^{3,-2})
\end{array}
\right..
\end{equation}
Therefore, the anti-symmetric SOC term $h_{IB}$ contains p-wave momentum functions and septet spin tensors.
The anti-symmetric SOC term $h_{IB}$ is parity-odd
\begin{equation}
P^{\dagger}h_{IB}(-\mathbf{k})P=-h_{IB}(\mathbf{k}),
\end{equation}
where $P$ stands for the inversion operation.

Let us denote the Green functions without inversion breaking term as $G_e^{(0)}(\mathbf{k},i\omega_n)$ and $G_h^{(0)}(\mathbf{k},i\omega_n)$. Given the fact that the energy scale of $h_{IB}$ is much smaller than other energy scale, e.g. $h_{SOC}\sim 1 eV$, we can choose the limit $0<\sqrt{2m/\mu}|C|\ll 1$ and thus
the Green functions with $h_{IB}$ can be expressed as
\begin{equation}
G_e(\mathbf{k},i\omega_n)=G_e^{(0)}(\mathbf{k},i\omega_n)+G_e^{(0)}(\mathbf{k},i\omega_n)h_{IB}(\mathbf{k})G_e^{(0)}(\mathbf{k},i\omega_n)+O(\sqrt{2m/\mu}|C|)^2
\end{equation}
\begin{equation}
\gamma G_h(\mathbf{k},i\omega_n)\gamma^{-1}=\gamma G_h^{(0)}(\mathbf{k},i\omega_n)\gamma^{-1}-\gamma G_h^{(0)}(\mathbf{k},i\omega_n)\gamma^{-1}h_{IB}(\mathbf{k})\gamma G_h^{(0)}(\mathbf{k},i\omega_n)\gamma^{-1}+O(\sqrt{2m/\mu}|C|)^2,
\end{equation}
where the latter uses the fact that $h_{IB}$ is time reversal invariant.
We can see, the first-order change of the Green functions given by $h_{IB}$ is parity-odd.
Since all zero-order terms in the linearied gap equation for singlet-quintet mixing are parity-even, $h_{IB}$ would not change the linearized gap equation for singlet-quintet mixing to the first order.
Therefore, it is reasonable for us to neglect the effect of small $h_{IB}$ to the singlet-quintet pairing mixing.
However, as discussed in Ref.\cite{PhysRevLett.116.177001}, $h_{IB}$ can mix the s-wave singlet channel with a p-wave septet channel belonging to $A_1$ irrep of $O_h$ group.
By considering that channel, the interaction Hamiltonian becomes
\begin{equation}
H_{int}=H_{int}^{(0)}+\frac{1}{2\mathcal{V}}\sum_{\mathbf{k},\mathbf{k}'}
\left[
V_2 c^{\dagger}_{\mathbf{k}}\left(\frac{a}{\sqrt{3}}\mathbf{k}\cdot\mathbf{V}\gamma\right)\left(c^{\dagger}_{-\mathbf{k}}\right)^T\left(c_{-\mathbf{k}'}\right)^T\left(\frac{a}{\sqrt{3}}\mathbf{k}'\cdot\mathbf{V}\gamma\right)^{\dagger}c_{\mathbf{k}'}
\right],
\end{equation}
where $H_{int}^{(0)}$ represents the original singlet-quintet interaction Hamiltonian (\ref{chosen_interaction_form}) and $\mathbf{V}=(V_x,V_y,V_z)$.
Thus, the gap function becomes
\begin{equation}
\label{paring_form_IB}
\Delta(\mathbf{k})
=\Delta^{(0)}(\mathbf{k})+\Delta_2 \frac{a}{\sqrt{3}}\mathbf{k}\cdot\mathbf{V}\gamma,
\end{equation}
where $\Delta^{(0)}(\mathbf{k})$ is the original singlet-quintet pairing form (\ref{paring_form}) and $\Delta_2$ is the order parameter of the p-wave septet channel.
Since this channel is parity-odd and has the similar form as $h_{IB}$, the mixing between this channel and s-wave singlet channel can exist for the first order of $h_{IB}$, which should be much smaller than singlet-quintet mixing that is controlled by $h_{SOC}$.

Next we treat the anti-symmetric SOC $h_{IB}$ and the p-wave quintet order parameter $\Delta_2$, which preserve the time-reversal symmetry\cite{PhysRevLett.116.177001,PhysRevB.96.144514}, as a perturbation, and exam its influence on the nodal-line superconductivity.
Due to the topological protection of nodal-line superconductivity, such small time-reversal and particle-hole invariant perturbation cannot directly gap out the nodal lines.
We find that this term can split one $N_w=\pm 2$  nodal line into two $N_w=\pm 1$ nodal lines since it breaks the inversion symmetry.
We exam the energy dispersion of the BdG Hamiltonian with the parameters chosen as $\tilde{\Delta}_0/|\mu|=1$,$\tilde{\Delta}_1/|\mu|=1.6$, $\tilde{\Delta}_2/|\mu|=0.1$ ,  $|2m| c_1=0.8$, $|2m| c_2=0.5$ and $\sqrt{2m/\mu} C=0.1$ to include the inversion breaking terms. The projection of bulk nodal lines (dark lines) onto (111) plane and the energy dispersion on (111) surface along $(11\bar{2})$ axis are shown in Fig.\ref{3half_pairing_c1neqc2_Inversion_Breaking} c and d, respectively.
Here $\tilde{\Delta}_2\equiv \Delta_2\sqrt{2 m \mu}a$.
Compared with bulk nodal line without inversion breaking term shown in Fig.\ref{3half_pairing_c1neqc2_Inversion_Breaking}a,
the previous one nodal line does split into two nodal lines as shown in Fig.\ref{3half_pairing_c1neqc2_Inversion_Breaking}c.
Compared with surface energy dispersion without inversion breaking term shown in Fig.\ref{3half_pairing_c1neqc2_Inversion_Breaking}b,
the previous one bulk touch point does split into two touching points as shown in Fig.\ref{3half_pairing_c1neqc2_Inversion_Breaking}d.
From these plots, we conclude that zero energy Majorana flat bands still exist for the inversion-breaking case as shown in Fig.\ref{3half_pairing_c1neqc2_Inversion_Breaking}d.
\begin{figure}[h]
\includegraphics[width=0.4\columnwidth]{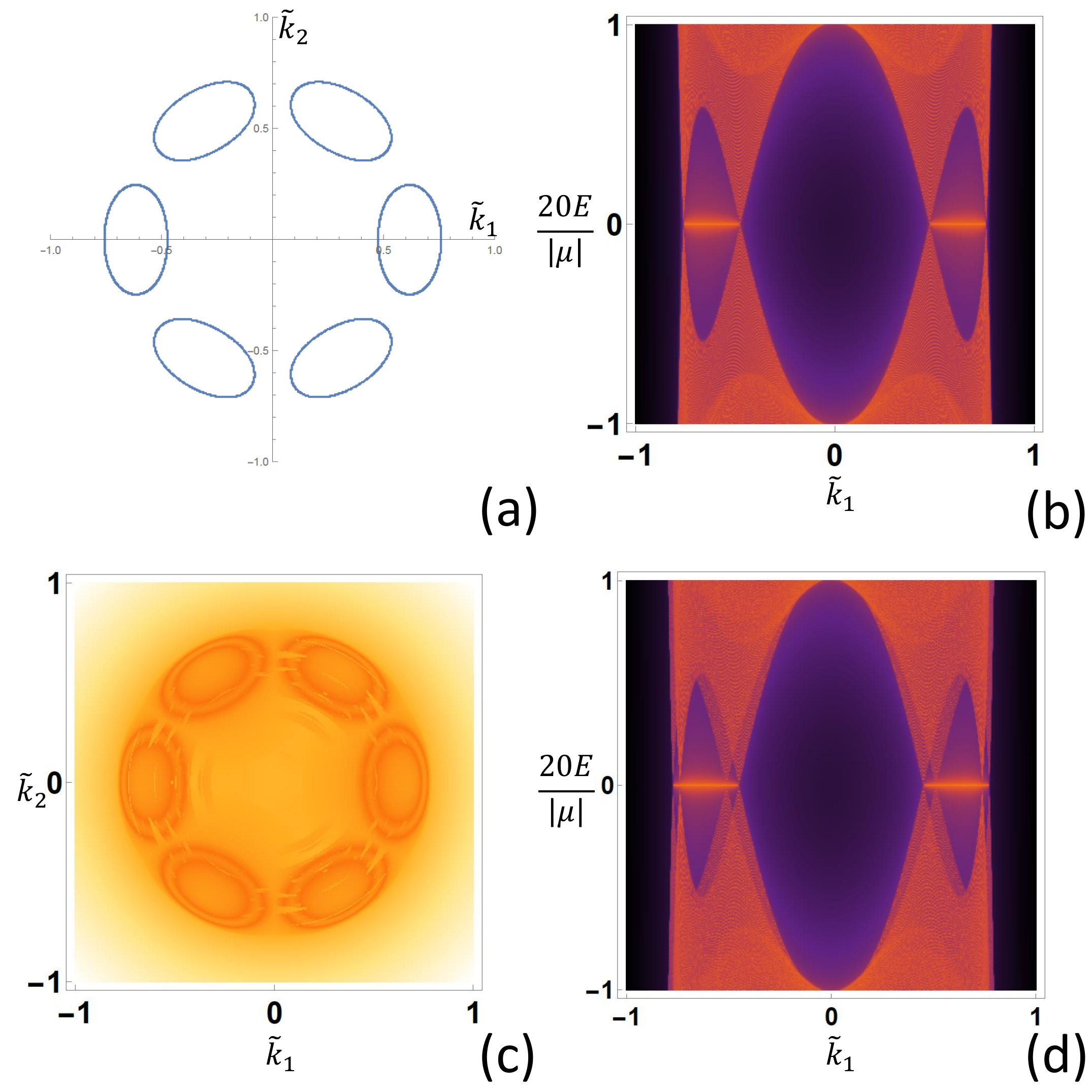}
\caption{
\label{3half_pairing_c1neqc2_Inversion_Breaking}
(a) and (b) show the projection of bulk nodal lines onto (111) plane and the energy dispersion on (111) surface along $(11\bar{2})$ axis, respectively, for $\tilde{\Delta}_0/|\mu|=1$,$\tilde{\Delta}_1/|\mu|=1.6$, $\tilde{\Delta}_2/|\mu|=0$,  $|2m| c_1=0.8$, $|2m| c_2=0.5$ and $\sqrt{2m/\mu} C=0$.
(c) and (d) show the projection of bulk nodal lines(dark lines) onto (111) plane and the energy dispersion on (111) surface along $(11\bar{2})$ axis, respectively, for $\tilde{\Delta}_0/|\mu|=1$,$\tilde{\Delta}_1/|\mu|=1.6$, $\tilde{\Delta}_2/|\mu|=0.1$,  $|2m| c_1=0.8$, $|2m| c_2=0.5$ and $\sqrt{2m/\mu} C=0.1$.
$\tilde{k}_{1,2}=k_{1,2}/\sqrt{2m\mu}$ are momentum along $(11\bar{2})$ and $(\bar{1}10)$ axes, respectively.
}
\end{figure}

\subsection{Nodal Superconductivity with Repulsive interaction in the quintet channel}
Another interesting situation occurs for the case that the interaction in the singlet channel is attractive $(\tilde{V}_0<0, \tilde{\lambda}_0>0)$, thus inducing the superconductivity, while that in the quintet channel is repulsive $(\tilde{V}_1>0, \tilde{\lambda}_1<0)$ within the energy cut-off $\epsilon_c$. This situation can occur when electron-phonon interaction dominates the singlet channel while repulsive Coulomb interaction dominates the quintet channel.
In this case, superconductivity can still exist and the singlet-quintet mixing can be solved by Eq.(\ref{linearized_gap_eqn_approx}). The expressions of the pairing ratio Eq.(\ref{Delta_ratio}) and transition temperature Eq.\ref{Expression_Tc} remains unchanged.

Since the quintet channel is repulsive $\tilde{\lambda}_1<0$,
it is necessary to require $(\lambda_0 y_{1}-\tilde{\lambda}_1 y_3)^2+4 \lambda_0 \tilde{\lambda}_1 y_{2}^2\geq 0$ for superconductivity to exist.
This requirement suggests the singlet channel would be strongly suppressed by the repulsive quintet channel interaction in the strong mixing limit. The discussion below will always assume this condition.

According to Eq. (\ref{Delta_ratio}), Eq. (\ref{nodal_condition_E+_without_lieqn}) and Eq. (\ref{nodal_condition_E-_without_lieqn}), the approximate nodal condition is present as follows.

(i)If $y_2>0$ which means $\tilde{\Delta}_0/\tilde{\Delta}_1<0$, nodal points cannot exist on $\xi_-$ Fermi surface.
In this case, the nodal condition for $\xi_+$ band (which is also the nodal condition for the whole system) is
\begin{equation}
\label{nodal_condition_E+_attractive_quintet}
1+2m Q_c > 0 \ \&\ k=\sqrt{2 m_+ \mu} \ \&\ \frac{\tilde{\lambda}_1}{\lambda_0}= \frac{f_+ y_1-y_2}{-f_+^2 y_2+f_+ y_3}\ \&\ \ y_2>0\ \&\ \  -\frac{y_2}{f_+}[f_+^2\left(\frac{f_+ y_1-y_2}{-f_+^2 y_2+f_+ y_3}\right)+1]\geq 0,
\end{equation}
where $f_+=\frac{1}{1+2m Q_c}\frac{|c_1|Q_1^2+|c_2| Q_2^2}{Q_c}$.

(ii)If $y_2<0$ which means $\tilde{\Delta}_0/\tilde{\Delta}_1>0$, nodal points cannot exist on $\xi_+$ Fermi surface.
In this case, the nodal condition for $\xi_-$ band (which is also the nodal condition for the whole system) is
\begin{equation}
\label{nodal_condition_E-_attractive_quintet}
1-2m Q_c > 0 \ \&\ k=\sqrt{2 m_- \mu} \ \& \ \frac{\tilde{\lambda}_1}{\lambda_0}= \frac{f_- y_1+y_2}{f_-^2 y_2+f_- y_3}\ \&\ \ y_2<0\ \&\ \ \frac{y_2}{f_-}[f_-^2\left(\frac{f_- y_1+y_2}{f_-^2 y_2+f_- y_3}\right)+1]\geq 0,
\end{equation}
where $f_-=\frac{1}{1-2 m Q_c}\frac{|c_1|Q_1^2+|c_2| Q_2^2}{Q_c}$.

\subsubsection{Regime I: Normal Band Structure}
When the interaction in the quintet channel is attractive,
nodal points in regime I (normal band structure) only exist on the $\xi_-$ Fermi surface, as discussed in Sec.\ref{Attractive_Regime_I_Nodal}.
For repulsive interaction in the quintet channel, the positive $y_2$ in regime I requires nodal points to only exist on the Fermi surface of the $\xi_+$ bands according to Eq. (\ref{nodal_condition_E+_attractive_quintet}).
To illustrate it, we choose $|2m c_2|=0.5$ and plot the phase diagram in the parameter space of  $|2m c_1|$ and $|\tilde{V}_1/V_0|$ with $0\leq |2m c_1|<1$ in Fig.\ref{3half_pairing_c1neqc2_Nodal_Line_repulsive_quintet_channel_Normal}a.
In this parameter region, the superconductivity exists for any $\tilde{\lambda}_1<0$ and $\lambda_0>0$.
While the system is gapped in the white region of Fig.\ref{3half_pairing_c1neqc2_Nodal_Line_repulsive_quintet_channel_Normal}a,
nodal lines exist on $\xi_+$ Fermi surface in the yellow region of Fig.\ref{3half_pairing_c1neqc2_Nodal_Line_repulsive_quintet_channel_Normal}a.
We find this situation (Fig.\ref{3half_pairing_c1neqc2_Nodal_Line_repulsive_quintet_channel_Normal}b)
is the same the case shown in Fig.2b of the main text (six loops centered about (001) axes or eight loops centered about (111) axes).
However, the six-loop(eight-loop) type nodal lines only exist in the left(right) yellow region of Fig.\ref{3half_pairing_c1neqc2_Nodal_Line_repulsive_quintet_channel_Normal}a.
The two yellow regions are disconnected since the system is gapped between the two dashed lines in Fig.\ref{3half_pairing_c1neqc2_Nodal_Line_repulsive_quintet_channel_Normal}a.
Therefore, no Lifshitz transition happens between two types of nodal lines in this case.
Moreover, the nodal lines have non-trivial 1d AIII topological invariant $(N_w=\pm 2)$ and can lead to surface Majorana flat bands.

\begin{figure}[h]
\includegraphics[width=0.5\columnwidth]{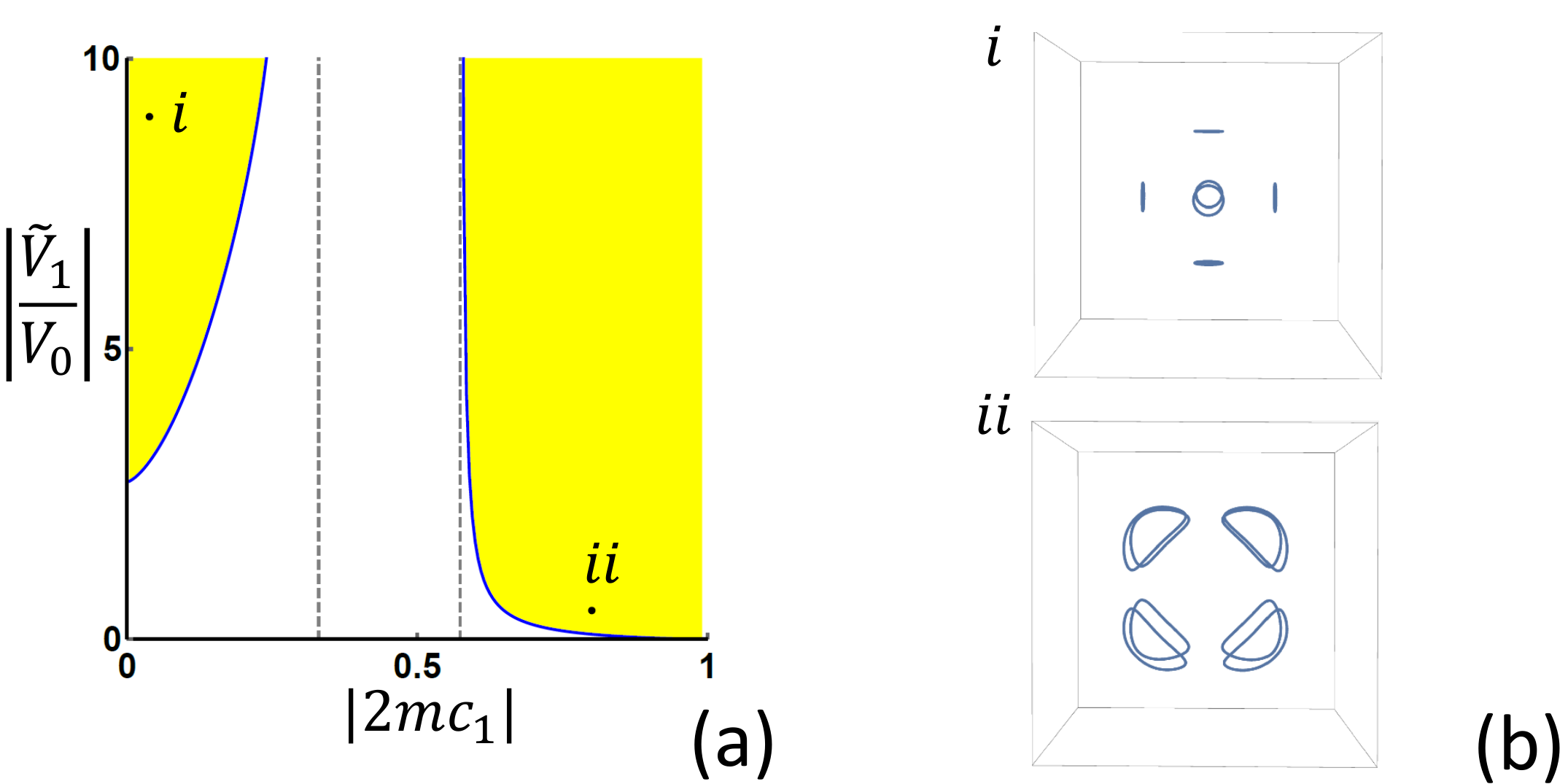}
\caption{
\label{3half_pairing_c1neqc2_Nodal_Line_repulsive_quintet_channel_Normal}
Note:$\tilde{V}_1/V_0=\tilde{\lambda}_1/\lambda_0$.
(a) shows the phase diagram in the parameter space $(|2m c_1|,|\tilde{V}_1/V_0|)$ for repulsive quintet channel $(\tilde{V}_1>0, \tilde{\lambda}_1<0)$,attractive singlet channel $(\tilde{V}_0<0, \tilde{\lambda}_0>0)$ and $2m|c_2|=-0.5$.
The system is gaped in the white region and nodal in the yellow region.
The yellow region extends to infinitely large $|\tilde{V}_1/V_0|$.
The two yellow regions are disconnected since the system is gapped between the two dashed lines.
(b) shows the bulk nodal structures for point $i: (|2mc_1|=0.04,|\tilde{V}_1/V_0|=9)$ and point $ii: (|2mc_1|=0.8,|\tilde{V}_1/V_0|=0.5)$ in (a).
}
\end{figure}

\subsubsection{Regime II: Inverted Band Structure}
It has been demonstrated in Sec.\ref{Attractive_Regime_II_Nodal} that nodal superconductivity cannot exist in regime II when the interaction in the quintet channel is attractive.
In contrast, we will demonstrate below that nodal lines are possible to appear on the Fermi surface of $\xi_-$ bands in regime II when the interaction in the quintet channel is repulsive.
Fig.\ref{3half_pairing_c1neqc2_Nodal_Line_repulsive_quintet_channel_Inverted}a reveals the phase diagram in the parameter space of $|2m c_1|$ and $|\tilde{V}_1/V_0|$ for $|2m c_2|=1.5$.
We notice the nodal superconductivity (yellow region in Fig.\ref{3half_pairing_c1neqc2_Nodal_Line_repulsive_quintet_channel_Inverted}a) can exist for a strong repulsive interaction when $|\tilde{V}_1/\tilde{V}_0|$ reaches $\sim 5$.
The form of the nodal lines (Fig.\ref{3half_pairing_c1neqc2_Nodal_Line_repulsive_quintet_channel_Inverted}b) is similar to that discussed in Fig. 2b of the main text.

\begin{figure}[h]
\includegraphics[width=0.5\columnwidth]{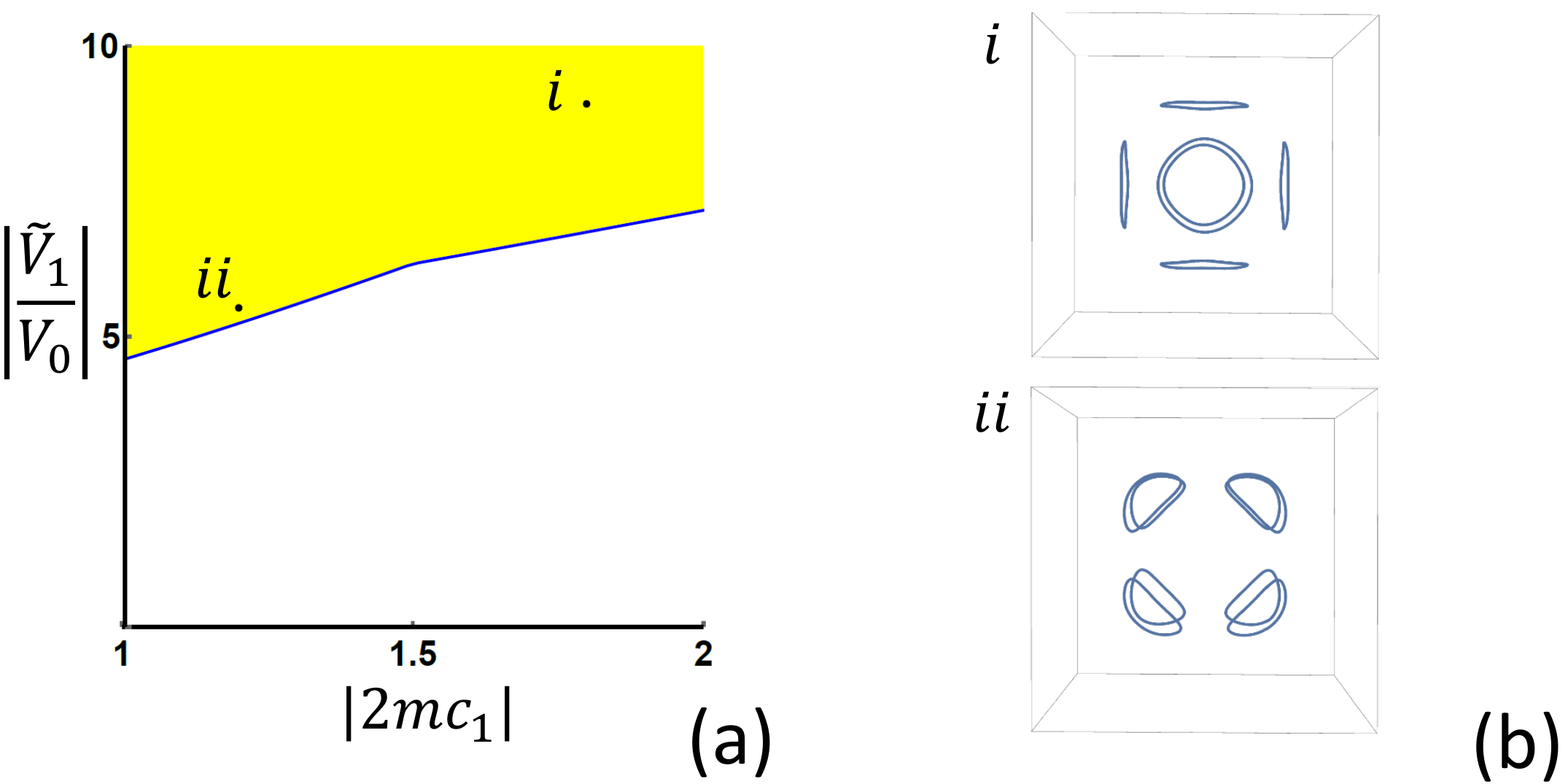}
\caption{
\label{3half_pairing_c1neqc2_Nodal_Line_repulsive_quintet_channel_Inverted}
Note:$\tilde{V}_1/V_0=\tilde{\lambda}_1/\lambda_0$.
(a) shows the phase diagram in the parameter space $(|2m c_1|,|\tilde{V}_1/V_0|)$ for repulsive quintet channel $(\tilde{V}_1>0, \tilde{\lambda}_1<0)$,attractive singlet channel $(\tilde{V}_0<0, \tilde{\lambda}_0>0)$ and $2m|c_2|=-1.5$.
The system is gaped in the white region and nodal in the yellow region.
The yellow region extends to infinitely large $|\tilde{V}_1/V_0|$.
(b) shows the bulk nodal structures for point $i: (|2mc_1|=1.8,|\tilde{V}_1/V_0|=9)$ and point $ii: (|2mc_1|=1.2,|\tilde{V}_1/V_0|=5.5)$ in (a).
}
\end{figure}

\subsubsection{Regime III: A Special Type of Inverted Band Structure with saddle point}

Repulsive quintet channel does not change the main result of Sec.\ref{Attractive_Regime_III_Nodal}: it is still possible to have nodal points on either of $\xi_{\pm}$ Fermi surfaces.
The phase diagram is shown in Fig.\ref{3half_pairing_c1neqc2_Nodal_Line_repulsive_quintet_channel_Intermediate} a ($0\leq |2m c_1|\lesssim 0.888$) and c ($0.888\lesssim |2m c_1|<1$).
Again, in this parameter region, the superconductivity exists for any $\tilde{\lambda}_1<0$ and $\lambda_0>0$.
When $0\leq |2m c_1|\lesssim 0.888$($0.888\lesssim |2m c_1|<1$), $y_2>0$($y_2<0$) and nodal lines exist on the $\xi_+$($\xi_-$) Fermi surface in the yellow region of Fig.\ref{3half_pairing_c1neqc2_Nodal_Line_repulsive_quintet_channel_Intermediate} a(c) according to Eq.\ref{nodal_condition_E+_attractive_quintet}(\ref{nodal_condition_E-_attractive_quintet}).
The dashed line $|2m c_1|\approx 0.888$ is(is close to) the asymptote of the phase boundary in Fig.\ref{3half_pairing_c1neqc2_Nodal_Line_repulsive_quintet_channel_Intermediate} a(c).
The nodal line distribution (Fig.\ref{3half_pairing_c1neqc2_Nodal_Line_repulsive_quintet_channel_Intermediate}  b and d)
is the same as Fig.2b of the main text.
Since the bottoms of the two nodal regions are far from each other, we split the phase diagram into two parts.

\begin{figure}[h]
\includegraphics[width=0.5\columnwidth]{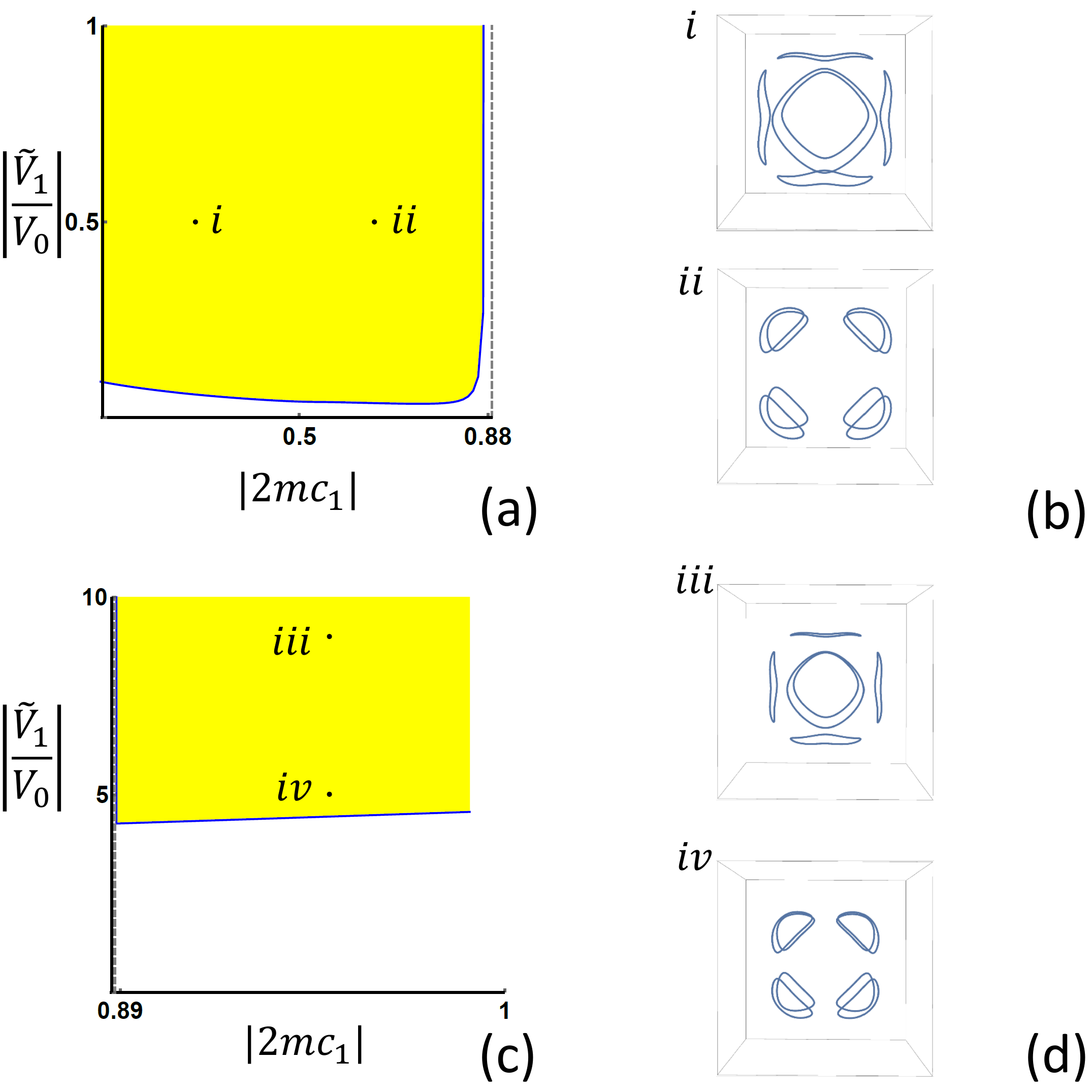}
\caption{
\label{3half_pairing_c1neqc2_Nodal_Line_repulsive_quintet_channel_Intermediate}
Note:$\tilde{V}_1/V_0=\tilde{\lambda}_1/\lambda_0$.
(a) and (c) show the phase diagram in the parameter space $(|2m c_1|,|\tilde{V}_1/V_0|)$ for repulsive quintet channel $(\tilde{V}_1>0, \tilde{\lambda}_1<0)$,attractive singlet channel $(\tilde{V}_0<0, \tilde{\lambda}_0>0)$, $\Lambda/\sqrt{2m\mu}=3$ and $2m|c_2|=-1.5$.
(a) and (c) are for $0\leq |2m c_1|\lesssim 0.888$ and $0.888\lesssim  |2m c_1|<1$, respectively.
The system is gaped in the white region and nodal in the yellow region.
The yellow region extends to infinitely large $|\tilde{V}_1/V_0|$.
The dashed line $|2m c_1|\approx 0.888$ is (or close to) the asymptote of the boundary of the yellow region in (a) and (c).
(b) shows the bulk nodal structures for point $i: (|2mc_1|=0.29,|\tilde{V}_1/V_0|=0.5)$ and point $ii: (|2mc_1|=0.65,|\tilde{V}_1/V_0|=0.5)$ in (a).
Nodal lines exist on $\xi_+$ Fermi surface.
(d) shows the bulk nodal structures for point $iii: (|2mc_1|=0.95,|\tilde{V}_1/V_0|=9)$ and point $iv: (|2mc_1|=0.95,|\tilde{V}_1/V_0|=5)$ in (c).
Nodal lines exist on $\xi_-$ Fermi surface.
}
\end{figure}

\subsection{Difference between our case and nodal superconductivity due to singlet-septet mixing}
Topological nodal-line superconductivity can also appear due to singlet-septet mixing and results in Majorana flat band (MFB) at the surface \cite{PhysRevB.96.094526}. We notice that in singlet-septet mixing, MFB regions at the surface are connected by Fermi arcs (See Fig.5(a) of Ref.\cite{PhysRevB.96.094526}).
This feature is absent for the MFB in our case (see Fig. 2(c) in the main text), due to the presence of inversion symmetry, as discussed below. Such different forms of MFB are possible to be observed experimentally and thus provides us an experimental signature to distinguish topological nodal-line superconductivity that originates from singlet-septet mixing or from singlet-quintet mixing.

As pointed out in Ref.\cite{PhysRevB.96.094526}, the Fermi arc is due to the non-trivial 1d AIII topological invariant defined for the mirror subspaces of the Hamiltonian along the direction perpendicular to the surface.
Without loss of generality, we consider the $(1\bar{1}0)$ mirror plane since other mirror planes can be related by symmetries.
The $(1\bar{1}0)$ mirror operation $M_{1\bar{1}0}$ on the BdG bases is represented as
\begin{equation}
M_{1\bar{1}0}=
-
\left(
\begin{matrix}
e^{- i \frac{J_x-J_y}{\sqrt{2}}\pi} & \\
 & [e^{- i \frac{J_x-J_y}{\sqrt{2}}\pi}]^*\\
\end{matrix}
\right)
=
-
\left(
\begin{matrix}
e^{- i \frac{J_x-J_y}{\sqrt{2}}\pi} & \\
 & e^{+ i \frac{J_x^*-J_y^*}{\sqrt{2}}\pi}\\
\end{matrix}
\right)\ .
\end{equation}
Since $M_{1\bar{1}0}$ is a unitary matrix, we can define a unitary matrix $U_M$ that diagonalizes $M_{1\bar{1}0}$:
\begin{equation}
U_M M_{1\bar{1}0} U_M^{\dagger}=
\left(
\begin{matrix}
i \mathds{1}_{4} & \\
 & -i \mathds{1}_{4}\\
\end{matrix}
\right)\ ,
\end{equation}
where $\mathds{1}_{n}$ is the $n\times n$ identity matrix.
$h_{BdG}(\mathbf{k}_{\perp})$ is invariant under the mirror operator $M_{1\bar{1}0} h_{BdG}(\mathbf{k}_{\perp}) M_{1\bar{1}0}^{\dagger}=h_{BdG}(\mathbf{k}_{\perp})$ with $\mathbf{k}_{\perp}$ being the momentum perpendicular to $(1\bar{1}0)$ direction $\mathbf{k}_{\perp}\cdot (1,-1,0)=0$.
Thus, $h_{BdG}(\mathbf{k}_{\perp})$ can be block diagonalized by $U_M$
\begin{equation}
U_M h_{BdG}(\mathbf{k}_{\perp})U_M^{\dagger}=
\left(
\begin{matrix}
h_+(\mathbf{k}_{\perp}) & \\
 & h_-(\mathbf{k}_{\perp})\\
\end{matrix}
\right)\ ,
\end{equation}
where $h_{\pm}(\mathbf{k}_{\perp})$ stands for the two mirror subspaces of $h_{BdG}(\mathbf{k}_{\perp})$ with mirror eigenvalues $\pm i$.
Since the time-reversal operation commutes with the mirror operation and all mirror eigen-values are purely imaginary, the time-reversal matrix can be block off-diagonalized by $U_M$
\begin{equation}
U_M\mathcal{T}U_M^T
=
\left(
\begin{matrix}
0 & \tilde{\gamma}^T\\
- \tilde{\gamma} & 0\\
\end{matrix}
\right)\ ,
\end{equation}
where $\tilde{\gamma}$ is a $4\times 4$ unitary matrix.
Therefore, the time-reversal symmetry gives
\begin{equation}
\tilde{\gamma}h_+^*(-\mathbf{k}_{\perp})\tilde{\gamma}^{\dagger}=h_-(\mathbf{k}_{\perp}).
\end{equation}
With $h_+(-\mathbf{k}_{\perp})=h_+(\mathbf{k}_{\perp})$ due to inversion symmetry, we have
\begin{equation}
\label{TRP_H_relation_Mirror}
\tilde{\gamma}h_+^*(\mathbf{k}_{\perp})\tilde{\gamma}^{\dagger}=h_-(\mathbf{k}_{\perp})\ .
\end{equation}
Since the chiral operation commutes with the mirror operation, the chiral operator can also be block diagonalized as
\begin{equation}
U_M\chi U_M^{\dagger}
=
\left(
\begin{matrix}
\tilde{\chi}_+ & \\
 & \tilde{\chi}_-\\
\end{matrix}
\right)\ ,
\end{equation}
where $\tilde{\chi}_{\pm}$ are $4\times 4$ unitary matrices.
Thus, both mirror subspaces have chiral symmetry, given by
\begin{equation}
\tilde{\chi}_{\pm} h_{\pm}(\mathbf{k}_{\perp})\tilde{\chi}^{\dagger}_{\pm}=-h_{\pm}(\mathbf{k}_{\perp}).
\end{equation}
As a result, $h_{\pm}$ can be transformed into an block off-diagonal form as
\begin{equation}
U_{\tilde{\chi}_{\pm}}h_{\pm}(\mathbf{k}_{\perp})U_{\tilde{\chi}_{\pm}}^{\dagger}=
\left(
\begin{matrix}
0 & D_{\pm}(\mathbf{k}_{\perp}) \\
D_{\pm}^{\dagger}(\mathbf{k}_{\perp}) & 0\\
\end{matrix}
\right)\ ,
\end{equation}
where $D_{\pm}(\mathbf{k}_{\perp})$ are $2\times 2$ matrices and $U_{\tilde{\chi}_{\pm}}$ are unitary matrices that diagonalize $\tilde{\chi}_{\pm}$
\begin{equation}
U_{\tilde{\chi}_{\pm}}\tilde{\chi}_{\pm}U_{\tilde{\chi}_{\pm}}^{\dagger}=
\left(
\begin{matrix}
-\mathds{1}_2 &  \\
 & \mathds{1}_2\\
\end{matrix}
\right)\ .
\end{equation}
The 1-d AIII topological invariant can be defined for both $h_{\pm}(\mathbf{k}_{\perp})$ according to Eq. (\ref{N_Arg_Q}) and Eq. (\ref{N_Arg_h}) as
\begin{equation}
\label{N_Arg_Dpm}
N_{w,\pm}=\frac{1}{2\pi }\int_{\mathcal{L}} d\mathbf{k}\cdot \mathbf{\nabla}_{\mathbf{k}}\text{Arg}[\text{Det}(D_{\pm}(\mathbf{k}))],
\end{equation}
where $\mathcal{L}$ is a closed/infinitely-long path on $(1\bar{1}0)$ plane along which $h_{\pm}(\mathbf{k}_{\perp})$ is gapped.
The total $N_w$ defined in Eq. (\ref{N_Arg_h}) can be decomposed into
\begin{equation}
N_w=N_{w,+}+N_{w,-}
\end{equation}
if the path $\mathcal{L}$ is on the mirror plane.
The above definition is slightly different from the corresponding one in Ref.\cite{PhysRevB.96.094526} due to the opposite sign in the definition of $N_-$.

Due to Eq. (\ref{TRP_H_relation_Mirror}), we have
\begin{equation}
\label{D_-D_+}
\left(
\begin{matrix}
0 & D_{-}(\mathbf{k}_{\perp}) \\
D_{-}^{\dagger}(\mathbf{k}_{\perp}) & 0\\
\end{matrix}
\right)
=
U_{\tilde{\chi}_-}h_{-}(\mathbf{k}_{\perp})U_{\tilde{\chi}_-}^{\dagger}
=
U_{\tilde{\chi}_-}\tilde{\gamma}U_{\tilde{\chi}_+}^T
\left(
\begin{matrix}
0 & D_{+}^*(\mathbf{k}_{\perp}) \\
D_{+}^{T}(\mathbf{k}_{\perp}) & 0\\
\end{matrix}
\right)
(U_{\tilde{\chi}_-}\tilde{\gamma}U_{\tilde{\chi}_+}^T )^{\dagger}
\ .
\end{equation}
Since $\mathcal{T}\chi^*=-\chi\mathcal{T}$, we have $-\tilde{\gamma}\tilde{\chi}_+^*=\tilde{\chi}_-\tilde{\gamma}$, which means $U_{\tilde{\chi}_-}\tilde{\gamma}U_{\tilde{\chi}_+}^T $ should take a block off-diagonal form as
\begin{equation}
\label{chit_gammat}
U_{\tilde{\chi}}\tilde{\gamma}U_{\tilde{\chi}}^T
=
\left(
\begin{matrix}
 &  \tilde{\gamma}_1\\
\tilde{\gamma}_2 & \\
\end{matrix}
\right)\ ,
\end{equation}
where $\tilde{\gamma}_{1,2}$ are $2\times 2$ unitary matrices.
Combining Eq. (\ref{D_-D_+}) with Eq. (\ref{chit_gammat}) leads to
$
D_{-}(\mathbf{k}_{\perp})=\tilde{\gamma}_1 D_{+}^T(\mathbf{k}_{\perp})\tilde{\gamma}_2^{\dagger}
$ and thus
\begin{equation}
\text{det}[D_{-}(\mathbf{k}_{\perp})]=\text{det}[\tilde{\gamma}_1\tilde{\gamma}_2^{\dagger}]\text{det}[D_{+}(\mathbf{k}_{\perp})]\Rightarrow \text{Arg}\{\text{det}[D_{-}(\mathbf{k}_{\perp})]\}=\text{Arg}\{\text{det}[\tilde{\gamma}_1\tilde{\gamma}_2^{\dagger}]\}+\text{Arg}\{\text{det}[D_{+}(\mathbf{k}_{\perp})]\}\
.
\end{equation}
Further with Eq. (\ref{N_Arg_Dpm}) and the fact that $\text{Arg}\{\text{det}[\tilde{\gamma}_1\tilde{\gamma}_2^{\dagger}]\}$ is k-independent, we arrive at
\begin{equation}
N_{\omega_-}=N_{\omega,+}.
\end{equation}
Therefore, we conclude that $N_w=N_{w,+}+N_{w,-}=2 N_{w,+}$ for path $\mathcal{L}$ on the mirror plane.

In Fig.2c of the main text, outside the MFB region,
we have $N_w=0$ for the infinite long path $\mathcal{L}$ along $(111)$ direction with specific $k_{1,2}$.
$N_w=0$ leads to $N_{w,-}=N_{w,+}=0$ if the path $\mathcal{L}$ is on a mirror plane.
Therefore, there is no Fermi arc at the edges of mirror planes outside the MFB region.
In our case, the time-invariant $A_1$ p-wave septet pairing, which also preserves mirror symmetry, is introduced as as a perturbation and thus will not induce any Fermi arc outside the MFB region.

\end{widetext}
\end{document}